\documentclass[useAMS,referee, usenatbib]{biom}

\def\bSig\mathbf{\Sigma}

\usepackage[figuresright]{rotating}
\usepackage{array}
\usepackage{float}
\usepackage{rotfloat}
\usepackage{multirow}
\usepackage{amsmath}
\usepackage{graphicx}
\usepackage{color}

\title[Dynamic Prediction of Competing Risk Events]{Dynamic Prediction of Competing Risk Events using Landmark Sub-distribution Hazard Model with Multiple Longitudinal Biomarkers}

\author{Cai Wu$^{1,2}$, 
Liang Li$^{2,*}$\email{LLi15@mdanderson.org}, and Ruosha Li$^{1}$ \\
$^{1}$Department of Biostatistics, The University of Texas Health Science Center at Houston, \\ Houston TX 77030, U.S.A. \\
$^{2}$Department of Biostatistics, The University of Texas MD Anderson Cancer Center, \\ Houston TX 77030, U.S.A.}

\begin{document}









\label{firstpage}


\begin{abstract}
The cause-specific cumulative incidence function (CIF) quantifies the subject-specific disease risk with competing risk outcome. With longitudinally collected biomarker data, it is of interest to dynamically update the predicted CIF by incorporating the most recent biomarker as well as the cumulating longitudinal history. Motivated by a longitudinal cohort study of chronic kidney disease, we propose a framework for dynamic prediction of end stage renal disease using multivariate longitudinal biomarkers, accounting for the competing risk of death. The proposed framework extends the landmark survival modeling to competing risks data, and implies that a distinct sub-distribution hazard regression model is defined at each landmark time. The model parameters, prediction horizon, longitudinal history and at-risk population are allowed to vary over the landmark time. When the measurement times of biomarkers are irregularly spaced, the predictor variable may not be observed at the time of prediction. Local polynomial is used to estimate the model parameters without explicitly imputing the predictor or modeling its longitudinal trajectory. The proposed model leads to simple interpretation of the regression coefficients and closed-form calculation of the predicted CIF. The estimation and prediction can be implemented through standard statistical software with tractable computation. We conducted simulations to evaluate the performance of the estimation procedure and predictive accuracy. The methodology is illustrated with data from the African American Study of Kidney Disease and Hypertension. \\
\end{abstract}

%



\maketitle


%

\section{Introduction}
\label{s:intro}

Patients with chronic kidney disease (CKD) are at increased risk of end stage renal disease (ESRD). Accurate prediction of the timing is of great importance in clinical research and practice to facilitate preparation for renal replacement therapy and individualize clinical decisions \citep{tangri2011predictive}. The typical ESRD risk equations are ``static'' prediction models in the sense that they are developed from survival regression models that relate the predictors at an earlier time point, such as baseline, to the time of ESRD \citep{echouffo2012risk, greene2017static}. Longitudinal data of those biomarkers between baseline and the terminal event are often available and potentially informative to the disease progression, but they are not used in prediction model development. 

In statistical literature, the prediction of the risk of clinical events using longitudinal data is often referred to as dynamic prediction in the sense that the prediction can be updated with accumulating longitiudinal data. Important fundamental work has been published in the last decade \citep{van2011dynamic, van2007dynamic, van2008dynamic, zheng2005partly, rizopoulos2011dynamic, proust2009development}. There are a number of challenges when this methodology is applied to the prediction of ESRD among CKD population. First, CKD patients have increased chance of mortality before reaching ESRD. Proper adjustment for competing risk is often needed in CKD studies \citep{noordzij2013we}. Second, previous literature has identified a large number of risk factors, including multiple biomarkers that are known to be causally associated with ESRD \citep{tangri2011predictive}. Put in the longitudinal context, it requires that the dynamic prediction model should accommodate multiple biomarkers with tractable computation. Some biomarkers, such as the estimated GFR, have diverse nonlinear progression trajectories \citep{li2012longitudinal, li2013within}. This feature could add to the complexity of the statistical analysis if modeling subject-specific longitudinal trajectories is needed. Third, it may take many years before a CKD patient reaches ESRD or death. The strength of association between biomarkers and the disease outcome may vary over time, leading to time-varying effects. Fourth, it is common that patients do not always follow a pre-specified clinical visits schedule. Even if the visit times are non-informative in the sense that they are not related to the health condition of the patients, the irregularly spaced and unsynchronized biomarker measurement times pose a challenge to the development of dynamic prediction model, as elucidated below.

On the topic of dynamic prediction with competing risks data, \citet{rue2017bayesian} and \citet{andrinopoulou2017combined} modeled the joint distribution of longitudinal data and competing risks with shared random effect models, and estimated the parameters with Markov chain Monte Carlo. However, when a large number of random effects are needed to accommodate multiple and possibly nonlinear longitudinal trajectories per subject, fitting the joint model is computationally infeasible \citep{hickey2016joint}. \citet{cortese2010competing}, \citet{nicolaie2013adynamic} and \citet{nicolaie2013bdynamic} studied the problem using an alternative, computationally simpler approach called landmark modeling \citep{van2011dynamic}. Motivated by the specific needs of CKD research, our proposed methodology in this paper is different from the statistical literature above in some important aspects. First, the typical landmark approach involves pre-specifying a number of landmark time points distributed over the follow-up period, creating a landmark dataset at each landmark time point that consists of at-risk subjects and their predictor variables and time-to-event, and fitting the model to the stacked landmark datasets \citep{van2011dynamic, nicolaie2013adynamic}. The predictor variable is not always measured at the landmark times due to irregularly spaced and unsynchronized measurement times. Imputing the unknown value by the last or closest measurement is difficult to apply because that measurement could be years apart and because the progression of CKD includes both chronic periods, when biomarkers change slowly, and acute episodes, when biomarkers change more quickly. Our proposed method does not require pre-specification of the landmark times, which is important given that there is currently no guideline in the literature on how to set the number and locations of the landmark times. It can also accommodate the irregularly spaced observational times without explicit imputation. Second, the proposed method estimates the time-varying model parameters semiparametrically without imposing a parametric shape \citep{nicolaie2013adynamic, van2011dynamic}. Lastly, our approach is embedded within the framework of Fine-Gray sub-distributional hazard model \citep{fine1999proportional}, while the previous works were based on cause-specific hazard model \citep{nicolaie2013adynamic}, pseudo-observations \citep{nicolaie2013bdynamic} and multi-state model \citep{cortese2010competing}. The Fine-Gray model imposes a parsimonious relationship between predictors and the cumulative incidence function (CIF) of ESRD without a separate model for death, which is difficult to establish due to the heterogeneity in the causes of death.


\section{The landmark dynamic prediction model for competing risks data}\label{sec:cmprsk_mod}

\subsection{The notation and data structure}

Let $T_{i}$ and $C_{i}$ be the time to the event of interest and time of censoring for subject $i$, and $\epsilon_{i}\in \{ 1,\ldots, K \}$ be the $K$ causes of the event. We observe the follow-up time $\tilde{T_{i}}=\textrm{min}(T_{i},C_{i})$, the censoring indicator $\Delta_{i}=1(T_{i}\le C_{i})$, and event type $\Delta_{i}\epsilon_{i}\in \{ 0,\ldots,K \}$. Without loss of generality,
we assume $K=2$ throughout this paper. In the context of the data application, event 1 denotes ESRD, the clinical event of interest, and event 2 denotes death, the competing event. Let $\boldsymbol{Y}_{i}=[ \boldsymbol{Y}_{i1},\boldsymbol{Y}_{i2},\dots,\boldsymbol{Y}_{iq} ]$ denote the $n_{i}\times q$ matrix of subject $i$, with $n_{i}$ repeated measurements for each of the $q$ covariates. This notation covers both time-dependent (longitudinal) and time-independent (baseline) covariates. The repeated measurements are made at time points
$\boldsymbol{t}_{i}=\{ t_{i1},t_{i2},\dots t_{in_{i}} \}$ ($t_{ij}<\tilde{T}_{i}$), which are not necessarily the same for all subjects. At any follow-up time $u$ ($u<\tilde{T}_{i}$), we denote by $\boldsymbol{\mathcal{H}}_{i}(u)$ the observed covariate process within a history window $[u-\tau_{2},u]$, where $\boldsymbol{\mathcal{H}}_{i}(u)=\{\boldsymbol{Y}_{i}(t_{ij}), t_{ij} ~|~ u-\tau_{2}\le t_{ij}\le u ;\ j=1,\ldots,n_{i} \}$. We observe independent and identically distributed training data $\mathcal{D}_{n}=\{\tilde{T}_{i},\Delta_{i}\epsilon_{i},\boldsymbol{Y}_{i}, \boldsymbol{t}_i,i=1,\ldots,n\}$, from which the dynamic prediction model is to be developed.  Our interest is to estimate, for a future individual in the same population as the training data, indexed by subscript $_{o}$, the probability of ESRD in the next $\tau_{1}$ years (called prediction horizon) given survival up to time $s$ and the covariate information in the history window: $\pi(\tau_{1}\vert s,\boldsymbol{\mathcal{H}}_{o}(s))=P(T_{o}\in(s,s+\tau_{1}],\epsilon_{o}=1\vert\tilde{T}_{o}>s,\boldsymbol{\mathcal{H}}_{o}(s))$. We assume that the distribution of $\boldsymbol{t}_{i}$ is non-informative in the sense that it is independent of $\boldsymbol{Y}_{i}$ and $T_i$. We assume independent censoring in the sense that $C$ is conditionally independent of $T$ and $\epsilon$ given the baseline covariates.

 In the following, we describe the construction of landmark dataset. We use notation $T(s) = T - s$ to denote the residual lifetime when a generic subject is at risk at time $s$. For subject $i$ in the development dataset, we define $T_{ij}=T_{i}-t_{ij}$ and $C_{ij}=C_{i}-t_{ij}$ as the subject-specific residual times to event and censoring, starting from $t_{ij}$. For prediction up to the horizon $\tau_{1},$ we can artificially censor the residual times at $\tau_{1}$, i.e., we observe $\tilde{T}_{ij}=\textrm{min}(T_{ij},C_{ij},\tau_{1})$ and the event indicator $\tilde{\delta}_{ij}=1\big(T_{ij}\le\textrm{min}(C_{ij},\tau_{1})\big)\times\epsilon_{i}$, where $1(\cdot)$ is the indicator function. The artificial censoring helps to reduce the chance of misspecifying certain model assumptions \citep{van2011dynamic}. Throughout this paper, we focus on modeling the relationship between $T(s)$ and $\boldsymbol{H}(s)$ at any landmark time $s$. In the model development dataset, $T(s)$ and $\boldsymbol{H}(s)$ are observed at the longitudinal measurement times $t_{ij}$ ($i=1,2,...,n$, $j=1,2,...,n_i$), leading to the observed outcome and predictor data $\{ \tilde{T}_{ij},\tilde{\delta}_{ij},\boldsymbol{\mathcal{H}}_{i}(t_{ij}) \}$. Therefore, we also call $t_{ij}$ as the landmark time as they are the starting time of the residual lifetime outcome. When making a prediction, the new prediction is often made when new measurements become available, i.e., at a new $t_{ij}$ of that subject. From this perspective, the prediction time is also called a landmark time.

\subsection{Sub-distribution hazard model with baseline covariates}

{ \color{black}

We first briefly review the sub-distribution hazard (SDH) model for competing risks \citep{fine1999proportional} with baseline covariates $\boldsymbol{X}$. For prediction, the quantity of interest is the cumulative incidence function (CIF) given $\boldsymbol{X}$: $\pi_{1}(t^*;\boldsymbol{X})=P(T\le t^*,\epsilon=1\vert\boldsymbol{X})$. Under Fine and Gray's formulation, this CIF is formulated as
\begin{equation}
P(T\le t^*,\epsilon=1\vert\boldsymbol{X}) = 1 - exp\left( -\int_0^{t^*} \lambda_1(t|\boldsymbol{X}) dt \right)
\label{eq:baseFG_CIF}
\end{equation}
with $\lambda_1(t|\boldsymbol{X}) = \lambda_{10}(t)\textrm{exp}(\boldsymbol{\alpha}^{T}\boldsymbol{X})$, where $\lambda_{10}(t)$ can be any non-negative function of time $t > 0$ and $\boldsymbol{\alpha}$ is a real vector. Fine and Gray further developed an interpretation for the $\lambda_1(t|\boldsymbol{X})$ function. They showed that it can be interpreted as a sub-distribution hazard, in the sense that $\lambda_{1}(t;\boldsymbol{X})=\textrm{lim}_{\Delta t \rightarrow 0}\dfrac{1}{\Delta t}P(t \le T \le t+\Delta t,\epsilon=1\vert\{T\ge t\}\cup\{T\le t\cap\epsilon\ne 1\},\boldsymbol{X})=-\dfrac{d\textrm{log}(1-\pi_{1}(t;\boldsymbol{X}))}{dt}$.
Such a definition can be viewed as the hazard function for an improper
random variable $1(\epsilon=1)\times T+1(\epsilon\ne1)\times\infty$. This interpretation also helps the development of an estimation procedure that is analogous to that of the Cox model. 

As far as prediction is of concern, characterizing the bilateral relationship between time to event $T$ ($\epsilon=1$) and the covariate vector $\boldsymbol{X}$ is all that is needed. The sub-distribution hazard value at a time $t$ ($t>0$) is not of direct relevance to this prediction. The sub-distribution hazard function $\lambda_1(t|\boldsymbol{X})$ serves as the internal machinery that helps the estimation of model (\ref{eq:baseFG_CIF}). This is a key observation that motivates the working model in the next subsection.  
} 

\subsection{Landmark proportional sub-distribution hazard model}

{ \color{black}
By extending model (\ref{eq:baseFG_CIF}) to the context of dynamic prediction, we propose the following landmark proportional SDH model at landmark time $s$:
\begin{equation}
P(T(s)\le t^*,\epsilon=1\vert \boldsymbol{\mathcal{H}}(s), T > s ) = 1 - exp\left( -\int_0^{t^*} \lambda_1(t| \boldsymbol{\mathcal{H}}(s), s ) dt \right).
\label{eq:working_model_generic}
\end{equation}
As the notation on the left hand side of (\ref{eq:working_model_generic}) suggests, at any given landmark time $s$, this model is specified for those subjects still at risk ($T > s$) at that time. If we treat the given $s$ as a new baseline, then this model is equivalent to a model (\ref{eq:baseFG_CIF}) specified for the residual life time $T(s)$ among the at-risk subjects at time $s$, given the predictor variables defined from $\boldsymbol{\mathcal{H}}(s)$. Since there are in theory infinitely many landmark time $s$, model (\ref{eq:working_model_generic}) is formulated under the working assumption that these models hold simultaneously. For a specific landmark dataset $\{\tilde{T}_{ij},\tilde{\delta}_{ij},\boldsymbol{\mathcal{H}}_{i}(t_{ij}),t_{ij};\ i=1,\ldots,n,j=1,\ldots,n_{i}\}$, this model implies that the bilateral relationship between each residual time to event $( \tilde{T}_{ij},\tilde{\delta}_{ij} )$ and the corresponding ``baseline'' predictor variables extracted from history $\boldsymbol{\mathcal{H}}_{i}(t_{ij})$ satisfies model (\ref{eq:working_model_generic}) at the corresponding landmark time $s = t_{ij}$. To fit this model, we define a working SDH function as:
\begin{equation}
\lambda_{1}(t^{*}\vert\boldsymbol{\mathcal{H}}_{i}(t_{ij}),t_{ij})=\lambda_{10}(t^{*},t_{ij})\textrm{exp}\Big(\boldsymbol{\beta}^{T}(t_{ij})\boldsymbol{\tilde{Y}}_{i}(t_{ij})\Big),\ t^{*}\in(0,\tau_{1}],\label{eq:working_model}
\end{equation}
where $\lambda_{10}(t^{*},t_{ij})$ is a bivariate smooth baseline SDH function, defined on the scale of the residual lifetime $t^{*}\in(0,\tau_{1}]$ and landmark time $t_{ij}$. We use $\boldsymbol{\tilde{Y}}_{i}(t_{ij})$ to denote the predictors at visit time $t_{ij}$, which are functions of the observed history $\boldsymbol{\mathcal{H}}_i(t_{ij})$. 
The time-varying coefficients $\boldsymbol{\beta}(.)$ are assumed to be smooth functions to allow the association to vary with the landmark time. For the bilateral relationship between $( \tilde{T}_{ij},\tilde{\delta}_{ij} )$ and $\boldsymbol{\tilde{Y}}_{i}(t_{ij})$, the corresponding value of the coefficient function is $\boldsymbol{\beta}(t_{ij})$. 

Our landmark dataset construction resembles that in the partly conditional model \citep{zheng2005partly}, which resets the follow-up time scale at each landmark time. From this perspective, the basic idea of the proposed methodology is more closely related to that model than the landmark model of \citet{van2007dynamic}. However, besides the accommodation of competing risk outcome, another difference between our approach and \citet{zheng2005partly} is that the time-varying coefficients are functions of the landmark time $t_{ij}$ instead of the derived follow-up time $t^{*}$. Therefore it differs from the usual time-varying coefficient model in survival analysis that is commonly used to deal with non-proportional hazards \citep{cai2003local}. With the artificial censoring at $\tau_{1}$, the covariate effect is more likely to be constant over $t^* \in (0, \tau_1)$ (but still vary with $t_{ij}$) and the proportional sub-distribution assumption is more likely to hold. \citep{liu2016robust}. 

Model (\ref{eq:working_model}) is called a ``working'' sub-distribution hazard function because it is used to facilitate the model fitting using the estimating equations developed by Fine and Gray (1999). While it implies that a subject's residual sub-distribution hazard at landmark time $s$ is $ \lambda_{1}(t^{*} = 0 \vert \boldsymbol{\mathcal{H}}(s), s) $, it does not imply that this subject's sub-distribution hazard at time $s + t^*$ ($t^* > 0$) given the history $\boldsymbol{\mathcal{H}}(s)$, is still given by (\ref{eq:working_model}). The hazard at time $s + t^*$ depends on $\boldsymbol{\mathcal{H}}(s + t^*)$. In general, the hazard at time $s + t^*$ conditional on $\boldsymbol{\mathcal{H}}(s)$ depends on both the hazard at time $s + t^*$ conditional on $\boldsymbol{\mathcal{H}}(s+t^*)$ and the conditional distribution of the paths of longitudinal covariates  $\boldsymbol{\tilde{Y}}(u)$ ($u \in (s, s+t^*)$) given $\boldsymbol{\mathcal{H}}(s)$. This is elucidated by the concept of consistency \citep{Jewell1993}. Like other landmark (or partly conditional) models, the proposed model has not been proven as a consistent prediction model. However, this working model can still be a useful prediction tool as long as (\ref{eq:working_model_generic}) provides a good approximation to the bilateral relationship between $( \tilde{T}_{ij},\tilde{\delta}_{ij} )$ and $\boldsymbol{\tilde{Y}}_{i}(t_{ij})$.}  

\section{Model estimation and dynamic prediction of the CIF }\label{sec:mod_est}

For estimation, we extend the kernel approach in \citet{li2017dynamic} to the competing risk context and  formalize the idea of borrowing information from lagging covariates \citep{andersen2003attenuation,cao2015analysis}. Assume that $\boldsymbol{\beta}(.)$ has a continuous second
derivative in a neighborhood of $s$, by local linear approximation, $\boldsymbol{\beta}(t_{ij})\approx\boldsymbol{\beta}(s)+\boldsymbol{\beta}'(s)(t_{ij}-s)$ for
subject-specific time points $t_{ij}$ around $s$. The landmark dataset $\mathcal{L}_{m}$ are clustered multivariate time-to-event
data with competing events, where the $n_{i}$ records from the same
subject are correlated. For clustered competing risk data \citep{zhou2012competing},
we define the counting process for event 1 as $N_{ij}(t^{*})=1(t_{ij}\le T_{i}\le t_{ij}+t^{*},\tilde{\delta}_{ij}=1)$
and the at-risk process $R_{ij}(t^{*})=1-N_{ij}(t^{*}-)=1(T_{i}>t_{ij}+t^{*})+1(t_{ij}\le T_{i}\le t_{ij}+t^{*},\tilde{\delta}_{ij}\ne1)$. Based on a local ``working independence'' partial likelihood function \citep{zhou2012competing}, for any given landmark point $s$, we
can estimate the parameters $\boldsymbol{\beta}(s)$ using a kernel-weighted
estimation equation, by borrowing biomarker measurements from the
neighboring time points, $\{ t_{ij}\in(s-h,s+h)\}$: 
\begin{equation}
\sum_{i=1}^{n}\sum_{j=1}^{n_{i}}K_{h}(t_{ij}-s)\int_{0}^{t^{*}}w_{ij}(t)\cdot\Big\{\boldsymbol{\tilde{Z}}_{ij}(1,t_{ij}-s)-\boldsymbol{\bar{Z}}(\boldsymbol{\beta}(s),t)\Big\}\cdot\textrm{d}N_{ij}(t).\label{eq:est_eq}
\end{equation}
 $K(\cdot)$ is a kernel function with bounded support on $[-1,1]$.
$K_{h}(x)=h^{-1}K(x/h)$ and $h$ is the bandwidth. $\boldsymbol{\tilde{Z}}_{ij}(1,t_{ij}-s)=\boldsymbol{\tilde{Y}}(t_{ij})\otimes(1,t_{ij}-s)$ with $\otimes$ denoting the Kronecker product. We have the notations $\boldsymbol{\bar{Z}}(\boldsymbol{\beta}(s),t)=\dfrac{\hat{\boldsymbol{S}}^{(1)}(\boldsymbol{\beta}(s),t)}{\hat{\boldsymbol{S}}^{(0)}(\boldsymbol{\beta}(s),t)}$
, and 
\begin{align}
\hat{\boldsymbol{S}}^{(r)}(\boldsymbol{\beta}(s),t)=&n^{-1}\sum_{l=1}^{n}\sum_{m=1}^{n_{i}}K_{h}(t_{lm}-s)w_{lm}(t)R_{lm}(t)\times\boldsymbol{\tilde{Z}}_{lm}(1,t_{lm}-s)^{\otimes r}\nonumber\\
&\times\textrm{exp}\Big(\boldsymbol{b}^{T}(s)\boldsymbol{\tilde{Z}}_{lm}(1,t_{lm}-s)\Big),
\end{align}
where $\boldsymbol{b}(s)=\{\boldsymbol{b}_{0}(s),\boldsymbol{b}_{1}(s)\}=\{\boldsymbol{\beta}(s),\boldsymbol{\beta'}(s)\}$,
$\tilde{\boldsymbol{Z}}^{\otimes0}=1$ and $\tilde{\boldsymbol{Z}}^{\otimes0}=\tilde{\boldsymbol{Z}}$.
The coefficient $\boldsymbol{\beta}(s)$ is estimated at each landmark
$s$ using $\boldsymbol{\hat{\beta}}(s)=\boldsymbol{\hat{b}}_{0}(s)$.
The variance of $\hat{\boldsymbol{\beta}}(s)$ can be estimated by bootstrap, which involves randomly sampling $n$ subjects from the original dataset with replacement, estimating the point estimator from each randomly sampled bootstrap dataset, and calculating the sample variance of the point estimators from all bootstrap datasets \citep{bootstrap}. The $w_{ij}(\cdot)$ in (\ref{eq:est_eq}) denotes
the inverse probability censoring weight for competing events, modified
from \citet{fine1999proportional}: 
\[
w_{ij}(t^{*})=1\big(C_{ij}\ge T_{ij}\wedge t^{*}\big)\frac{G(t^{*}\vert s)}{G\big(T_{ij}\wedge t^{*}\vert s\big)},
\]
where $G(t^{*}\vert s)=P(C_{ij}\ge t^{*}\vert s)$ is the censoring
distribution of the residual censoring time at landmark $s$, and
$\wedge$ denotes the minimum of the two values. We use a kernel-weighted
Kaplan-Meier estimator for the residual censoring distribution, estimated from the residual time to censoring around $s$:
\[
\widehat{G}(t^{*}\vert s)=\prod_{\zeta\in\Omega,\zeta\le t}\Big\{1-\frac{\sum_{l}K_{h}(t_{lm}-s)\cdot1(\tilde{C}_{lm}=\zeta,\tilde{\delta}_{lm}=0)}{\sum_{l}K_{h}(t_{lm}-s)\cdot1(\tilde{C}_{lm}\ge\zeta)}\Big\}.
\]
Once we obtain the estimates of $\boldsymbol{\beta}(s)$, the baseline
cumulative SDH function at time $s$ can be estimated by plugging
in $\hat{\boldsymbol{\beta}}(s)$:
\[
\hat{\Lambda}_{10}(t^{*},s)=\dfrac{1}{n}\sum_{i=1}^{n}\sum_{j=1}^{n_{i}}K_{h}(t_{ij}-s)\int_{0}^{t^{*}}\frac{1}{\hat{\boldsymbol{S}}^{(0)}(\hat{\boldsymbol{\beta}}(s),t)}\hat{w}_{ij}(t)\textrm{d}N_{ij}(t).
\]
The conditional CIF for any future subject
$_{o}$ can be estimated as 
\begin{align}
\hat{\pi}_{1}(t^{*}\vert s,\boldsymbol{\mathcal{H}}_{o}(s)) &= \hat{P}(s<T_{o}\le s+t^{*},\epsilon_{o}=1\vert\tilde{T}_{o}>s,\boldsymbol{\mathcal{H}}_{o}(s))\nonumber \\
&= 1-\textrm{exp}\Big(-\hat{\Lambda}_{10}(t^{*},s)\times\textrm{exp}\Big(\hat{\boldsymbol{\beta}}^{T}(s)\boldsymbol{\tilde{Y}}_{o}(s)\Big)\Big).
\end{align}

\section{Quantifying the dynamic predictive accuracy}\label{sec:prederr}

In this section, we study two predictive accuracy measures, the time-dependent receiver operating characteristic (ROC) curve, in particular the area under the ROC curve (AUC); and
the Brier score (BS). In the dynamic prediction framework, the time-dependent
predictive accuracy measures are functions of two time scales, the
landmark time $s$ and the prediction horizon $\tau_{1}.$ The following procedure for estimating sensitivity, specificity, and BS were modified from the non-parametric kernel-weighted approach of \citet{wu2017quantifying} for competing risk data. 

\subsection{The dynamic time-dependent ROC curve and AUC}

At any landmark time $s$, we want to evaluate how well the
risk score, i.e., the estimated CIF, discriminates between subjects with the event of interest in the window $(s,s+\tau_{1}]$ versus those without. For any at-risk subject at time $s$ who experiences the main event within the time interval $(s,s+\tau_{1}]$, that occurrence is defined as a case: $D^{+}(s,\tau_{1})=\{i: s<T_{i}\le s+\tau_{1},\epsilon_{i}=1\}$. When a subject is event-free at $s+\tau_{1}$, that occurrence is defined as a
control: $D^{-}(s,\tau_{1})=\{i: T_{i}>s+\tau_{1}\}$. An alternative definition
for a control is to use the complementary set $\bar{D}^{+}(s,\tau_{1})=\{i: (s<T_{i}\le s+\tau_{1},\epsilon_{i}\ne1)\cup (T_{i}>s+\tau_{1})\}$, including subjects who experience a competing event within the time interval $(s,s+\tau_{1}]$ or remain event-free at $s+\tau_{1}$. To illustrate the ideas, we present the estimators for the former in this subsection. A similar extension can be made for the latter. For simplicity, we use the notation $U(\tau_{1}\vert s)$ to denote the individual predicted CIF (i.e., the risk score). Given a threshold value $c\in(0,1)$ , the time-dependent sensitivity and specificity functions are defined as $Se(c,s,\tau_{1})=P\Big(U(\tau_{1}\vert s)>c\vert D^{+}(s,\tau_{1})\Big)$
and $Sp(c,s,\tau_{1})=P\Big(U(\tau_{1}\vert s)\le c\vert D^{-}(s,\tau_{1})\Big).$ The estimators of sensitivity and specificity are
\begin{align*}
\widehat{Se}(c,s,\tau_{1}) & =\frac{\sum_{i\in\Re_{s}}\hat{W}_{1i}^{dyn}\cdot1(U_{i}(\tau_{1}\vert s)>c)}{\sum_{i\in\Re_{s}}\hat{W}_{1i}^{dyn}}\\
\widehat{Sp}(c,s,\tau_{1}) & =\frac{\sum_{i\in\Re_{s}}(1-\sum_{k=1}^{K}\hat{W}_{ki}^{dyn})\cdot1(U_{i}(\tau_{1}\vert s)\le c)}{\sum_{i\in\Re_{s}}(1-\sum_{k=1}^{K}\hat{W}_{ki}^{dyn})},
\end{align*}
where $W_{1i}^{dyn}=P\Big(T_{i}(s)\in(0,\tau_{1}],\epsilon_{i}=1\vert\tilde{T_{i}}(s),\delta_{i},U_{i}\Big)=1(\tilde{\delta}_{ij}=0)\cdot\frac{F_{1}(\tau_{1}\vert U_{i},s)-F_{1}(\tilde{T}_{i}(s)\vert U_{i},s)}{S(\tilde{T}_{i}(s)\vert U_{i},s)}+1(\tilde{\delta}_{ij}=1)$, $T_{i}(s)=T_{i}-s$, $\tilde{T_{i}}(s)=\tilde{T}_{i}-s$ and $U_{i}$ is short for $U_{i}(\tau_1\vert s)$. $\Re_s$ is risk set within the neighborhood of $s$ which includes the most recent record at $t_{ij}$ for each subject $i$ $\{i: \tilde{T}_{i}\ge s, \vert t_{ij}-s\vert \le \vert t_{ij^{'}}-s\vert, \forall j^{'}=1,2,\ldots,n_{i},t_{ij}\in (s-h,s+h)\}$. $F_1(x\vert U_{i}, s)=P(T_{i}(s)\le x, \epsilon_{i}=1\vert U_{i},s)$ and $S(x \vert U_{i},s)=P(T_i(s)\ge x \vert U_i,s)$.

For estimating the conditional probability weight $W_{1i}^{dyn}$, we treat the at-risk data set at landmark $s$ as the new baseline data set.
The time-dependent ROC curve is a plot of sensitivity $Se(c,s,\tau_{1})$
over 1-specificity $1-Sp(c,s,\tau_{1})$, i.e., for $x\in[0,1]$, $R\widehat{OC}(x,s,\tau_{1})=\widehat{Se}(\widehat{Sp}^{-1}(1-x,s,\tau_{1}),s,\tau_{1})$.
The AUC is estimated as $\widehat{AUC}(s,\tau_{1})=\int_{0}^{1}\widehat{ROC}(x,s,\tau_{1})\textrm{d}x$.

\subsection{The dynamic time-dependent Brier score}

The time-dependent BS under the dynamic competing risk framework
is defined as $BS(\tau_{1},s)=E\Big(1(s<T\le s+\tau_{1},\epsilon=1)-U(\tau_{1}\vert s)\big\vert T>s\Big)^{2}$, where $1(\cdot)$ is the indicator function.
Applying the weight $W_{1i}^{dyn}$, the BS can be estimated
as
\[
\widehat{BS}(\tau_{1},s)=\frac{1}{n_{s}}\sum_{i=1}^{n_{s}}\Big(\hat{W}_{1i}^{dyn}\times(1-U(\tau_{1}\vert s))^{2}+(1-\hat{W}_{1i}^{dyn})\times(0-U(\tau_{1}\vert s))^{2}\Big),
\]
where $n_{s}$ is the number of subjects at risk at landmark time $s$. 

The $AUC$ and $BS$ assess different aspects of the predictive model. $AUC$ evaluates
the discrimination between a case and a control, and $BS$ quantifies
the deviance of the predicted probability from the observed data. A model with
perfect discrimination will have $AUC=1$, while $AUC$ close to 0.5
indicates poor discrimination that resembles a random guess. $BS$ is
a prediction error metric, with smaller values indicating better prediction.

\section{Simulation}\label{sec:simu}

The simulation in this section mainly evaluates the prediction accuracy of the proposed model. A separate simulation, which evaluates the estimation of model parameters and bandwidth selection under the assumptions of the working model, is presented in the Web Appendix B. Similar to other studies evaluating the prediction accuracy of landmark models \citep{maziarz2017longitudinal}, we simulated longitudinal and competing risks data from a joint frailty model with shared random effects \citep{elashoff2008joint}. Details of the data generation process are described in the Web Appendix A. The data generating model included a baseline covariate and three longitudinal biomarkers. We considered two scenarios: (S1) the longitudinal biomarkers are non-informative for survival in the sense that their effects on both time-to-event outcome are zero, and (S2) the longitudinal biomarkers are informative in the sense that they have non-zero regression coefficients on both time-to-event outcomes. The incremental contribution of the longitudinal biomarkers to the prediction accuracy is expected to be zero under S1 and non-zero under S2. 


Table \ref{tab:simu_PE} presents the predictive accuracy of the proposed model under both S1 and S2. The full model (M1) includes both the longitudinal biomarkers and the baseline covariate; the null model (M0) only includes the baseline covariate. Since the data were simulated from a joint frailty model \citep{elashoff2008joint}, the proposed landmark SDH model worked under misspecification. However, regardless of whether the data generating model matches the fitting model, the predictive performance can always be evaluated. We considered both discrimination and calibration measures in Table \ref{tab:simu_PE}. For discrimination, we reported the true positive (TP) fraction and false positive (FP) fraction at a given threshold value, and the AUC as a global discrimination summary. For calibration, we used the Brier score. The predictive accuracy measures were evaluated at three landmark times $s=3,5,7$ with prediction horizon $\tau_{1}=1,3$. For each simulation, the proposed model was fit to a simulated training data set and the predictive accuracy measures were calculated from another simulated validation dataset from the same distribution.  When all the longitudinal biomarkers are non-informative, the predictive accuracy measures of the full model and the null model are very similar. When the three longitudinal biomarkers are informative, including the longitudinal biomarkers in the prediction model substantially improves both discrimination and calibration. 

\begin{sidewaystable}[ht]
\caption{The means (EST) and empirical standard deviations (ESD) of estimated predictive accuracy metrics comparing the full model with longitudinal biomarkers ($M_{1}$) and the null model with only the baseline covariate but without longitudinal biomarkers ($M_{0}$) in the simulation. Prediction horizon $\tau_{1}=3$. S1: non-informative longitudinal biomarkers. S2: informative longitudinal biomarkers. AUC: area under the ROC curve 
    comparing the group experiencing the event of interest with those who experienced competing events or event-free. $TP(c):$ true positive fraction at threshold $c$. $FP(c):$
    false positive fraction at threshold $c$. BS: Brier score. Sample size $n=500$.}
  \label{tab:simu_PE}
  \centering{}%
  \begin{tabular}{cccccccccccccc}
    \hline 
    &  &  & \multicolumn{2}{c}{$AUC$} &  & \multicolumn{2}{c}{$TP(0.25)$} &  & \multicolumn{2}{c}{$FP(0.25)$} &  & \multicolumn{2}{c}{$BS$}\tabularnewline
    \cline{4-5} \cline{7-8} \cline{10-11} \cline{13-14} 
    &  &  & $M_{1}$ & $M_{0}$  &  & $M_{1}$ & $M_{0}$  &  & $M_{1}$ & $M_{0}$  &  & $M_{1}$ & $M_{0}$ \tabularnewline
    \hline 
    \multirow{6}{*}{S1} & \multirow{2}{*}{$s=1$} & $EST$ & 0.703 & 0.707 &  & 0.514 & 0.512 &  & 0.272 & 0.290 &  & 0.161 & 0.165\tabularnewline
    &  & $ESD$ & 0.031 & 0.030 &  & 0.088 & 0.209 &  & 0.064 & 0.173 &  & 0.011 & 0.012\tabularnewline
    \cline{2-14} 
    & \multirow{2}{*}{$s=3$} & $EST$ & 0.691 & 0.698 &  & 0.726 & 0.675 &  & 0.529 & 0.499 &  & 0.199 & 0.203\tabularnewline
    &  & $ESD$ & 0.034 & 0.032 &  & 0.092 & 0.223 &  & 0.099 & 0.237 &  & 0.011 & 0.015\tabularnewline
    \cline{2-14} 
    & \multirow{2}{*}{$s=5$} & $EST$ & 0.660 & 0.676 &  & 0.751 & 0.705 &  & 0.614 & 0.594 &  & 0.211 & 0.223\tabularnewline
    &  & $ESD$ & 0.050 & 0.049 &  & 0.125 & 0.285 &  & 0.133 & 0.309 &  & 0.014 & 0.027\tabularnewline
    \hline 
    \multirow{6}{*}{S2} & \multirow{2}{*}{$s=1$} & $EST$ & 0.894 & 0.611 &  & 0.595 & 0.176 &  & 0.091 & 0.126 &  & 0.082 & 0.126\tabularnewline
    &  & $ESD$ & 0.031 & 0.061 &  & 0.117 & 0.217 &  & 0.027 & 0.175 &  & 0.034 & 0.077\tabularnewline
    \cline{2-14} 
    & \multirow{2}{*}{$s=3$} & $EST$ & 0.882 & 0.578 &  & 0.738 & 0.477 &  & 0.240 & 0.452 &  & 0.142 & 0.205\tabularnewline
    &  & $ESD$ & 0.030 & 0.059 &  & 0.132 & 0.345 &  & 0.122 & 0.342 &  & 0.064 & 0.070\tabularnewline
    \cline{2-14} 
    & \multirow{2}{*}{$s=5$} & $EST$ & 0.884 & 0.531 &  & 0.831 & 0.553 &  & 0.350 & 0.552 &  & 0.162 & 0.242\tabularnewline
    &  & $ESD$ & 0.030 & 0.058 &  & 0.082 & 0.400 &  & 0.113 & 0.392 &  & 0.035 & 0.052\tabularnewline
    \hline 
  \end{tabular}
\end{sidewaystable}

Under S2, the estimated regression parameters of the proposed SDH model are plotted in Web Figure 1, which shows that the effects of the three longitudinal biomarkers are notably different from zero and are in the right direction suggested by the data generating model. In contrast, the estimated regression parameters of the proposed SDH model are close to zero when data were generated under S1. 

We conducted additional simulation to compare the predicted and true conditional risks at individual level. The true conditional risk of an individual (indexed by subscript $o$, at landmark time $s$, conditional on biomarker history $\boldsymbol{\mathcal{H}}(s)$, and with prediction horizon $\tau_1$) is defined as $\pi(\tau_{1}\vert s, \boldsymbol{\mathcal{H}}_{o}(s))=P(T_{o}\in(s,s+\tau_{1}],\epsilon_{o}=1\vert\tilde{T}_{o}>s,\boldsymbol{\mathcal{H}}_{o}(s))$. Since the true conditional risk varies by subject, landmark time and history, and does not have a tractable analytical expression, we calculated it empirically at 9 representative landmark time by history combinations (Web Table 1) as follows. We simulated data using the procedure in Web Appendix A but with one informative longitudinal biomarker $Y_2$ (the effect of $Y_1$, $Y_3$, and $X$ on the time-to-event outcome were all set to zero). The true CIF for the event of interest in the next $\tau_1=3$ years were obtained empirically as the proportion of subjects with that event within $(s,s+\tau_1]$ give survival up to time $s$, among those with nearly identical $Y_2$. For illustration, we chose three target $Y_2$ values: $m = 0$, $2$ and $4$ in Web Table 1. Subjects with marker values within $\pm 0.05$ of the target value were counted in the denominator of the proportion calculation. To ensure that there were enough subjects in the denominator, we simulated a very large dataset ($n=1,000,000$) without censoring. We restricted to the case of a single informative biomarker ($Y_{2}$) because it is less feasible to match subjects on multiple biomarkers. The conditional CIFs for hypothetical subjects with marker value $0,2,4$ at landmark times $1,3,5$ were estimated from 500 Monte Carlo repetitions. The average estimated CIF (EST), empirical standard deviation (ESD), percent bias (\% Bias) and mean squared error (MSE) are presented in Web Table 1. Being a working model, the proposed semi-parametric landmark SDH model worked under mis-specification in this simulation. However, the result suggests that the estimated CIF has little bias and the MSE is low. It may indicate that the proposed model was flexible enough to provide approximate the data well at multiple landmark times, despite that the simulated data do not exactly satisfy the working assumptions. Unlike the simulation results in Table \ref{tab:simu_PE}, the results in Web Table 1 pertain to the quality of predictions at individual level.

The proposed landmark SDH model is a working model and it is not yet clear whether there exists a joint distirbution of longitudinal and competing risk data such that the model holds at all landmark times. This is a well known difficulty with landmark dynamic prediction models in general \citep{van2011dynamic, li2017dynamic} and it is not specific to our landmark model, though limited progress has been made in problems without competing risks \citep{zheng2005partly, DataGenerationPaper, JRSSCpaper}. Due to this difficulty, researchers often evaluate the numerical performance of the landmark models using data simulated from the joint model with shared random effects  \citep{maziarz2017longitudinal, huang2016two}. This is also the simulation strategy that we chose. When the data generation model and the analysis model do not match, the estimated model parameters are difficult to evaluate and interpret, but prediction accuracy can still be assessed. We have designed scenarios S1 and S2 to demonstrate that incorporating informative longitudinal biomarkers improves predictive accuracy. This qualitative conclusion is unlikely to be invalidated by the magnitude of model misspecification. We designed three longitudinal biomarkers with complicated trajectories to demonstrate that the proposed model works in situations where joint modeling approaches may be difficult to apply.

\section{Application to the AASK data}\label{sec:app}

The AASK study included $1,094$ African Americans of age 18 to 70 years who were diagnosed with hypertensive renal disease and had baseline eGFRs 
between $20-65 mL/min/1.73m^{2}$ \citep{wright2002effect}. Subjects were followed every 6 months, with up to 12 years of longitudinal data collected at each visit. By the end of the study, 318 (29\%) individuals developed ESRD, the event of research interest, and 176 (16\%) died before ESRD. The median time to ESRD was 4.3 years and the median time to death was 5.2 years. We chose clinically relevant prediction horizons of $\tau_{1}=1$ or $3$ years and illustrated the dynamic prediction at years 3, 5, and 7.  The key longitudinal biomarker is eGFR (estimated Glomerular Filteration Rate). Our previous publication demonstrated that this biomarker have diverse and possibly nonlinear individual progression patterns \citep{li2012longitudinal}. In addition, some CKD patients experienced acute kidney injury (AKI) during the follow-up, which may cause substantial short term variation in the eGFR (e.g., see Figure 2 of \citet{li2017dynamic}). The number of repeated measurements for eGFR ranged from 3 to 30, with over 50\% of individuals providing 17 or more measurements. In addition to the current value of eGFR at a clinical visit, we derived the rate of change in eGFR (linear eGFR slope) during the history window of $\tau_{2}=3$ years, because the eGFR slope is often used by clinicians to characterize the speed of progression in CKD \citep{Schluchter2001}. The estimation of eGFR slope followed the approach in our recent paper \citep{li2017dynamic}. Additional biomarkers included longitudinal measurements of serum albumin (Alb), urine protein to creatinine ratio (UP/Cr), serum phosphorus (Phos) and urine potassium (Upot). These biomarkers were considered because they have known biological association with disease progression and have been used in other CKD risk equations \citep{tangri2011predictive}. Also included in the prediction model were age at the time of prediction and an indicator of any hospitalization during the previous year.

\begin{figure}[ht]
  \begin{centering}
    \includegraphics[scale=0.7]{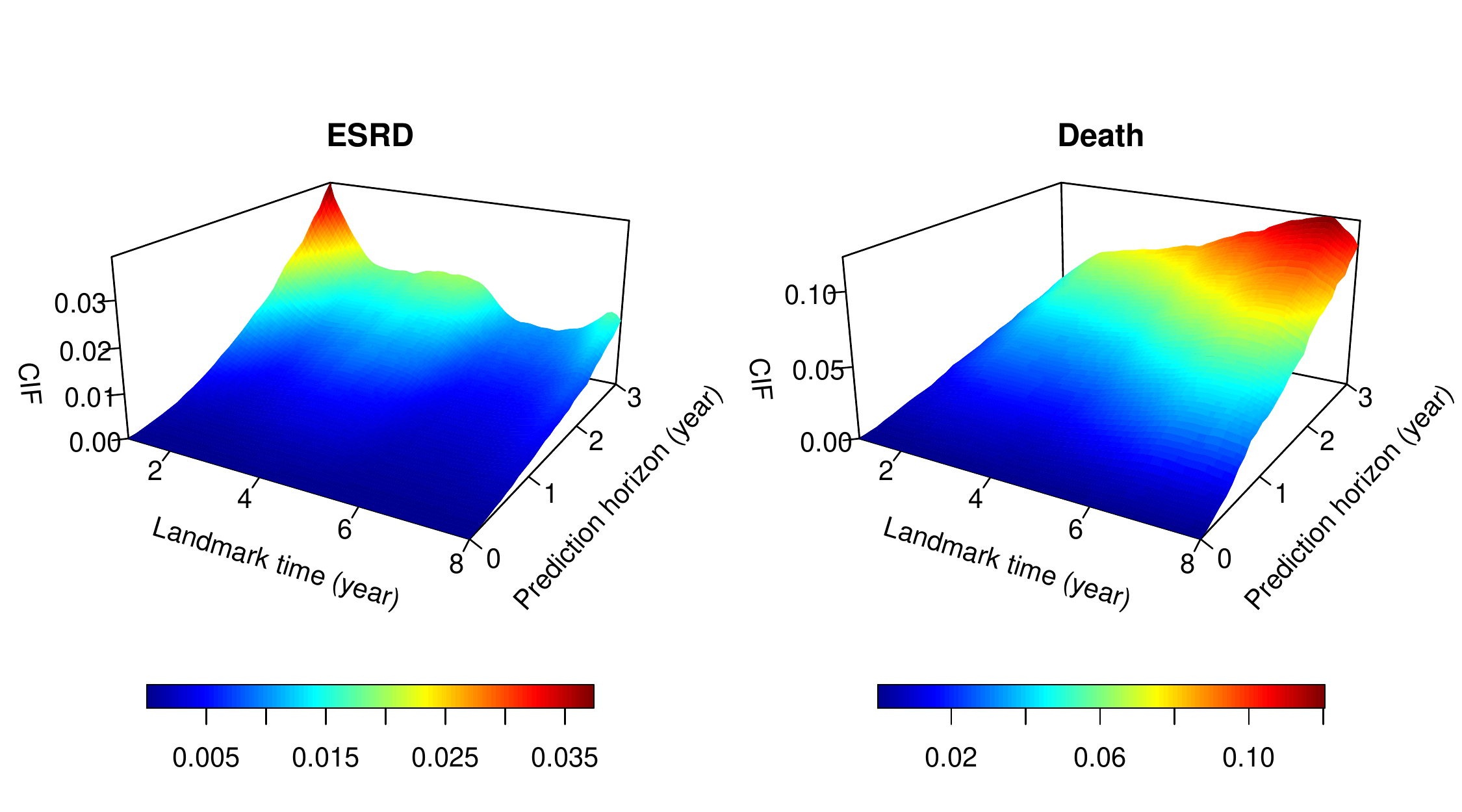}
    \par\end{centering}
  \caption{Estimated surface of the cumulative incidence function over the landmark time and prediction horizon. This shows an examplary population with age = 55, eGFR = 45 $ml/min/1.73m^{2}$, eGFR.slope = 0, UP/Cr = 0.3 $g/g$, albumin = 4 $g/dL$, and hospitalization within the past year. }
  \label{fig:surface_plot}
\end{figure}

For the competing events of ESRD and death, we fit landmark SDH models separately using the same set of candidate predictors (Web Figure 2). The eGFR, its rate of change, and log UP/Cr were significantly associated with time to ESRD but not with time to death. In contrast, age, Alb and hospitalization are risk factors related to death. This indicates that the progression to ESRD and death may be related to different pathological processes, which justifies the proposal of modeling the competing events separately rather than as a composite outcome. After removal of the non-significant covariates, the final model for ESRD included eGFR, eGFR.slope, log UP/Cr and Phos, and the final model for death included age, Alb, log-Upot and hospitalization (Web Figure 3). We conducted bandwidth selection using 5-fold cross-validation. Predictive accuracy metrics were evaluated in the cross-validation dataset, and they were robust to different bandwidths (up to 3 digits after the decimal point). Therefore, we used the bandwidth of $h = 1.5$ in the final model, which provided a relatively smooth curve for the log-SDH ratio curve.The surface plots of the CIF for ESRD and death are illustrated in Figure \ref{fig:surface_plot}. 

\begin{figure}[ht]
  \centering{}\includegraphics[scale=0.75]{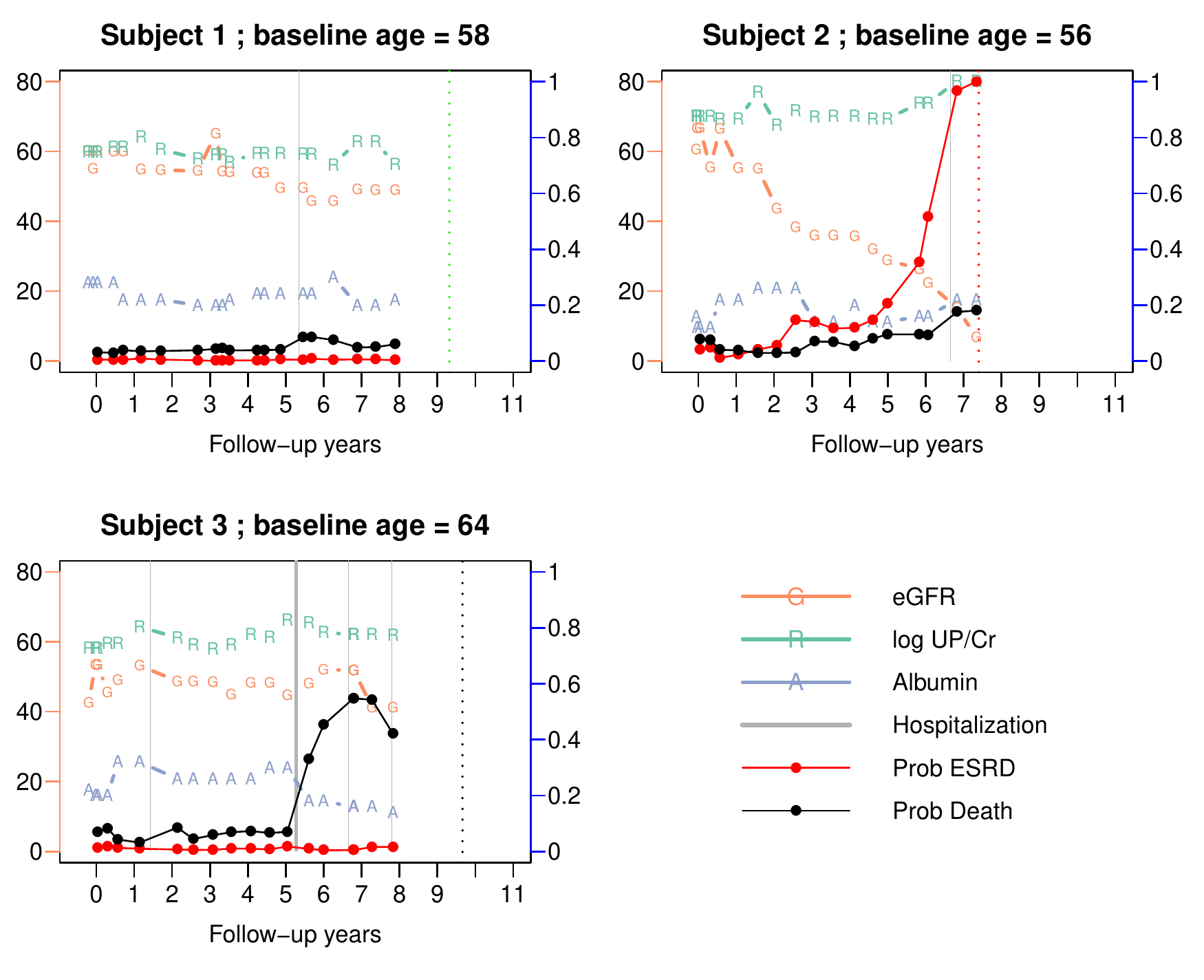}\caption{Individual risk predictions for three selected subjects: subject 1
    was censored (dotted vertical green line), subject 2 had ESRD (dotted
    vertical red line) and subject 3 died (dotted vertical black line). Three biomarkers are plotted over time: ``G'' is eGFR
    $(ml/min/1.73m^{2})$, ``R'' is log-urine protein-to-creatinine
    ratio ($g/g$), and ``A'' is albumin($g/dL$). The connected
    red dots are predicted probabilities of ESRD within a horizon of $\tau_{1}=3$ years. The gray vertical bars represent episodes of hospitalization, with the two vertical borders being admission and discharge dates. The connected black dots are the predicted probability of death within $\tau_{1}=3$ years. The y-axis to the left is the scale of eGFR, and the y-axis to the right is the scale of predicted probabilities (0 to 1). The other two biomarkers, log-UP/CR and albumin, are re-scaled to be displayed in the same plot with eGFR but their respective scales are not shown. The dynamic predicted probabilities of ESRD are calculated using the
    dynamic SDH model with four predictors: eGFR, eGFR slope in the past
    three years, log-UP/CR and phosphorus. The dynamic
    predicted probabilities of death are calculated using the dynamic SDH
    model with four predictors: current age, serum albumin, any
    hospitalization within the past year, and log urine potassium. }
  \label{fig:dynpred1}
\end{figure}

Figure \ref{fig:dynpred1} presents the longitudinal profiles and individual dynamic predictions from three AASK subjects: subject 1 was event-free by the end of the study, subject 2 experienced ESRD after 7.5 years, and subject 3 died after 9.7 years. We demonstrated the biomarker values with real-time predicted 3-year probabilities of ESRD and death. The risk prediction was dynamically updated at each new clinical visit. Subject 1 demonstrated stable disease. Subject 2 demonstrated persistent decline in eGFR and notable increase in proteinuria (log-UP/Cr), which led to drastic increase in the risk of ESRD after year 5. In contrast, the risk of death for subject 2 increased moderately, which may be explained by the Alb level and hospitalization around year 7. For subject 3, the relatively stable eGFR and log-Up/Cr also stablized the subject's susceptibility to ESRD, but the frequent hospitalization and decreasing Alb level were associated with increased risk of death, possibly due to other co-morbidity. We did not estimate the model after year 8 because the number of observed clinical events was relatively small near the end of the follow-up. 

\begin{sidewaysfigure}[ht]
  \centering{}\includegraphics[scale=0.75]{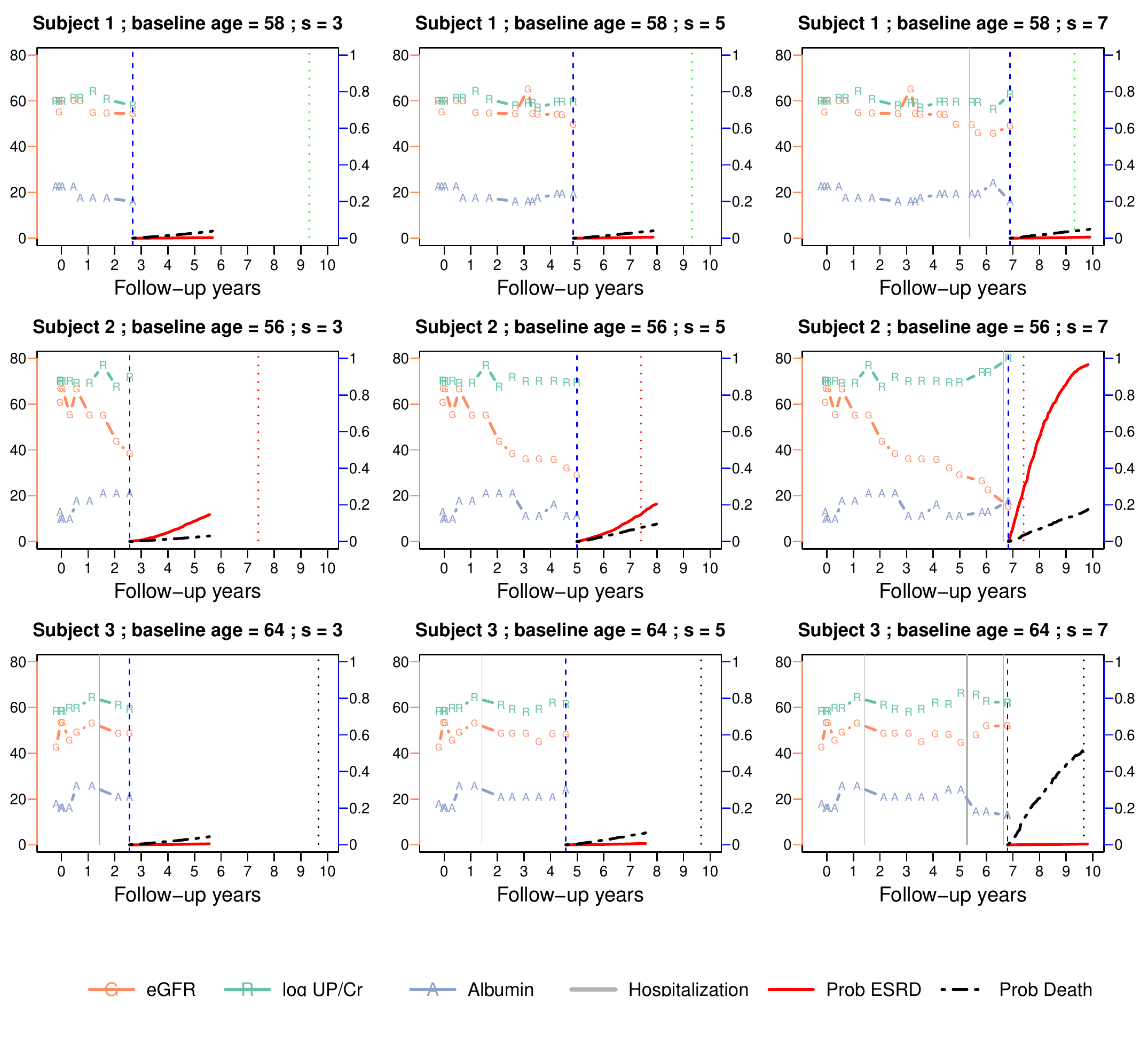}\caption{Individual dynamic predicted CIF for the three selected subjects in Figure \ref{fig:dynpred1}. Each row in the panel represents
    one subject, the three columns are the predictions made
    at landmark years $s=3,5,7$. The prediction is made at the vertical blue dashed lines. The predicted CIFs up to $\tau_{1}=3$
    years are plotted for the event of ESRD (red curve) and death (black
    curve). Symbols in the figure are similar to Figure \ref{fig:dynpred1}.}
  \label{fig:dynpred2}
\end{sidewaysfigure}

Figure \ref{fig:dynpred2} presents the profiles of the same three patients but with their dynamic CIFs (up to $\tau_1 = 3$ years) at landmark times $s=3,5,7$ years. For subject 1, the predicted CIFs for both ESRD and death were flat. In contrast, the predicted CIF of ESRD for subject 2 started to increase after year 5 and the increase became very prominent by year 7. This was likely caused by a combination of deteriorated renal function (eGFR) and proteinuria (log-UP/Cr). This patient reached ESRD shortly after year 7. The predicted CIF of ESRD for subject 3 stayed flat, but the CIF of death increased at year 7, after frequent hospitalization. This subject eventually died at year
9.6 without ESRD. 

In Table \ref{tab:AASK_PE}, we summarized the predictive accuracy of the landmark SDH models for prediction horizons $\tau_{1}=1,3$ at three landmark years $s=3,5,7$. The model for ESRD achieved good discrimination with AUCs between 0.93-0.96. When we used the cutoff value of 0.05, the sensitivity (TP) and specificity (1-FP) could be controlled within 0.80-0.90 range under all scenarios. The prediction accuracy metrics were similar with different prediction horizons. In contrast, the model for death discriminated no better than a random guess, resulting in AUCs around 0.5; the prediction errors were also at least twice as large as those from predicting ESRD. More importantly, the AUCs from the proposed model improved in comparison with previous studies where AUCs were around 0.8 and always less than 0.9 \citep{li2017dynamic,maziarz2017longitudinal}. One possible explanation is that these previous studies treat ``time to ESRD or death'' as a composite outcome. This introduces noise and diminishes the predictive accuracy because all-cause death is difficult to predict with the selected biomarkers, which are prognostically specific to renal disease. The ROC curves for predicting ESRD were plotted in Web Figure 4. 

\begin{sidewaystable}[ht]
  \begin{centering}
    \caption{Measures of predictive accuracy from the landmark SDH model for ESRD and death in the analysis of AASK data. The
      estimates were obtained at three landmark years, $s=3,5,7$, with prediction
      horizon $\tau_{1}=1,3$ years. AUC: area under the ROC curve. $TP(c)$: true positive rate at threshold $c$; $FP(c)$:
      false positive rate at threshold $c$ ; thresholds $c$ are selected
      to be 0.05 for ESRD and 0.01 for death. BS: Brier score. }
    \label{tab:AASK_PE}
    \par\end{centering}
  \centering{}%
  \begin{tabular}{ccccccccccccc}
    \hline 
    &  & \multicolumn{2}{c}{$AUC$} &  & \multicolumn{2}{c}{$TP(c)$} &  & \multicolumn{2}{c}{$FP(c)$} &  & \multicolumn{2}{c}{$BS$}\tabularnewline
    \cline{3-4} \cline{6-7} \cline{9-10} \cline{12-13} 
    &  & ESRD & Death &  & ESRD & Death &  & ESRD & Death &  & ESRD & Death\tabularnewline
    \hline 
    \multirow{3}{*}{$\tau_{1}=1$} & $s=3$ & 0.957 & 0.545 &  & 0.925 & 0.568 &  & 0.075  & 0.539  &  & 0.024  & 0.191\tabularnewline
    & $s=5$ & 0.925 & 0.547 &  & 0.885  & 0.585 &  & 0.100  & 0.596 &  & 0.026  & 0.043\tabularnewline
    & $s=7$ & 0.965 & 0.584 &  & 0.832  & 0.692 &  & 0.099  & 0.675  &  & 0.021  & 0.035 \tabularnewline
    \hline 
    \multirow{3}{*}{$\tau_{1}=3$} & $s=3$ & 0.944 & 0.558 &  & 0.876 & 0.468 &  & 0.119  & 0.378 &  & 0.054  & 0.119 \tabularnewline
    & $s=5$ & 0.943 & 0.520 &  & 0.852 & 0.498 &  & 0.140  & 0.549 &  & 0.052 & 0.093\tabularnewline
    & $s=7$ & 0.957 & 0.492 &  & 0.863 & 0.343 &  & 0.131  & 0.405  &  & 0.048 & 0.110 \tabularnewline
    \hline 
  \end{tabular}
\end{sidewaystable}

\section{Discussion}\label{sec:disc}

For CKD patients, estimating the time to ESRD is crucial for the timely treatment management. Dynamic prediction is an attractive tool for this purpose, because it is adaptive to the changing health condition and prognostic history of the patient. It enables real-time monitoring of patient risk. In this paper, we develop novel methodology for dynamic prediction of ESRD among the CKD patients, and we overcome a number of analytical hurdles, including competing events of death, irregularly spaced clinical visit times, multiple biomarkers with complicated longitudinal trajectories, time-varying at-risk population, and time-varying covariate-outcome association. Our proposed methodology is flexible, because the model parameters are estimated semi-parametrically. Hence, it can effectively mitigate the risk of model misspecification. This feature is very important for dynamic prediction models, because, as explained in Section \ref{sec:simu}, the landmark dynamic prediction model is a working model and needs to provide adequate approximation to the data at all landmark times. Another advantage of the proposed methodology is that it is computationally simple, and it can be implemented through standard statistical software for competing risks analysis, regardless of how many longitudinal biomarkers are included as predictors. In this paper, the estimation process was accomplished with the available R function $coxph()$ after translating the competing events into a counting process \citep{geskus2011cause}. We believe that the simplicity in computation makes the proposed methodology attractive for various practical situations, including applications with large datasets, a large number of biomarkers with complicated longitudinal trajectories, and other longitudinal prognostic information that cannot be easily modeled at an individual-level (e.g., hospitalization episodes and medication history).  

Our kernel-based estimation approach relies on the assumption that the clinical visit times are non-informative. Future work is needed to study dynamic prediction when the frequency of clinical visits is related to the health condition of the patients. The predictors in our proposed model framework include pre-specified features extracted from the data history. Automatic extraction of predictive features from the longitudinal history is another topic that we will pursue in future research.




\section*{Supplementary Materials}
Web Appendices, Web Figures, and R code are available with this paper at the Biometrics website on Wiley Online Library.

\section*{Acknowledgements}
The authors declare no potential conflicts of interest with respect to the research, authorship and publication of this article. This research was supported by grants from the U.S. National Institutes of Health (P30CA016672, U01DK103225, R01DK118079).

\bibliographystyle{biom}

\appendix


\label{lastpage}

\end{document}


\begin{center}
\Large \textbf{Web-based Supplementary Materials for \\ ``Dynamic Prediction of Competing Risk Events using Landmark Sub-distribution Hazard Model with Multiple Longitudinal Biomarkers'' \\ 
by Cai Wu, Liang Li, and Ruosha Li}
\end{center}

\normalsize

\section*{\textbf{Web Appendix A}: Data Generation Procedure for Simulation}

The simulation results are presented in main text of the paper. This section presents details of the data generation procedure and parameter settings. The longitudinal processes are generated from equation (1) below. We simulated a total of $n$ subjects with independent and identically distributed data for each simulation run. 
\begin{align}
Y_{i1}(t_{ij}) & =m_{i1}(t_{ij})+\epsilon_{i1}(t_{ij})=b_{i01}+b_{i11}\cdot t_{ij}+\epsilon_{i1}(t_{ij})\nonumber \\
Y_{i2}(t_{ij}) & =m_{i2}(t_{ij})+\epsilon_{i2}(t_{ij})=b_{i02}+b_{i12}\cdot t_{ij}^{3}+\epsilon_{i2}(t_{ij}) \label{eq:yiq-1} \\
\textrm{logit} & \{P(Y_{i3}(t_{ij})=1)\}=m_{i3}(t_{ij})=b_{i03}+b_{i13}\cdot t_{ij} ,  \nonumber
\end{align}

For both non-informative biomarker effect (S1) and informative biomarker effect (S2), the data were simulated according to the joint frailty model of longitudinal biomarkers and the competing risk event times (Elashoff et al, 2008). It includes the longitudinal sub-model and the following survival sub-model ($k=1,2$):
\begin{equation}
\lambda_{k}(t)=\lambda_{k0}(t)\textrm{exp}\{\gamma_{k}X_{i}+\sum_{q=1}^{3}\beta_{kq}m_{iq}(t)+v_{k}u_{i}\}.
\label{eq:JM_frailty}
\end{equation}

The baseline hazard for the time-to-event outcome follows Weibull distribution with scale and shape parameters of (0.02, 2.3) and (0.01, 2.4) for event 1 and event 2 respectively. The longitudinal sub-model includes three longitudinal biomarkers. The first one $Y_{i1}(.)$ is a continuous biomarker with a linear mean trajectory $m_{i1}(.)$. The second biomarker $Y_{i2}(.)$ has a nonlinear subject-specific mean trajectory. The third biomarker is binary with a logit-linear mean trajectory. For the first two biomarkers, $\epsilon_{i1}(.)$ and $\epsilon_{i2}(.)$ are random noises with $N(0,0.5^2)$ distribution. Each biomarker's longitudinal trajectory is characterized by two random effects, denoted by $\boldsymbol{b}_{ip}=(b_{i0p},b_{i1p})^T$ ($p=1,2,3$). In the case of a linear trajectory, such as the first biomarker, they represent the subject-specific random intercept and slope. We let $\boldsymbol{b}_{i}=(\boldsymbol{b}_{i1}^T,\boldsymbol{b}_{i2}^T,\boldsymbol{b}_{i3}^T)^{T}\sim MVN(\boldsymbol{\Omega},\boldsymbol{D})$, where $\boldsymbol{\Omega}=(2.8,-0.14,2.1,0.01,-1,0.3)$ denote the population mean. The covariance matrix $\boldsymbol{D}$
can be decomposed into $\boldsymbol{D}=diag(\boldsymbol{\sigma}_{q})\times\boldsymbol{R}\times diag(\boldsymbol{\sigma}_{q})$,
where the diagonal matrix $diag(\boldsymbol{\sigma}_{q})$ includes
elements $\boldsymbol{\sigma}_{q}=(\sigma_{01},\sigma_{11},\sigma_{02},\sigma_{12},\sigma_{03},\sigma_{13})=(0.9,0.1,0.9,0.005,0.9,0.1)$ and correlation matrix $\boldsymbol{R}$.   
\begin{equation*}
\mathbf{R} = 
\begin{pmatrix}
    1 & 0.26 & -0.5 & -0.3 & -0.5 & -0.3 \\
     & 1 & -0.65 & -0.3 & -0.5 & -0.3 \\
     &  & 1 & 0.35 & 0.5 & 0.3 \\
     &  &  & 1 & 0.5 & 0.3\\
     &  &  &  & 1 & 0.3 \\
     &  &  &  &  & 1
\end{pmatrix}.
\end{equation*}

In the survival sub-model, $u_{i}$ is the frailty term accounting for the correlation between two competing events, and the parameter $v_{1}$ is set to 1 to ensure identifiability. We let $u_{i}\sim N(0,\sigma_{u}^2)$
where $\sigma_{u}=0.5$. For S1, $\{\beta_{1q}\}$ and $\{\beta_{2q}\}$ are all set to be zero. For S2, we set $\{\beta_{1q}; q=1,2,3\}=(-1.2,0.3,1.5)$ and $\{\beta_{2q}; q=1,2,3\}=(-0.2,0.05,0.6)$. For both S1 and S2, the sub-model includes one baseline covariate $X_{i}\sim N(0.5,0.5^2)$ with regression coefficient $\gamma_1 = -1.5$ and $\gamma_2 = -1$. The censoring times are generated from a mixture of uniform distribution $\eta_{1}\textrm{Unif}(0,3)+\eta_{2}\textrm{Unif}(3,6)+\eta_{3}\textrm{Unif}(6,9)+\eta_{4}\textrm{Unif}(9,12)$, where the mixing probabilities $\eta_{1}$ to $\eta_{4}$ ($\sum_{i=1}^4{\eta_i}=1$) are chosen to control the censoring rate at approximately 25\%. For example, they equal to $(0.1, 0.1, 0.2, 0.6)$ for the simulation with informative biomarker and $(0.1, 0.1, 0.1, 0.7)$ for the simulation with non-informative biomarker. See the description of these two simulation scenarios below.

The random intercept and random slope (time effect) are assumed to be positively correlated for each biomarker. We allow $\boldsymbol{Y}_{i1}$ and $\boldsymbol{Y}_{i2}$ to have mild negative correlation, and $\boldsymbol{Y}_{i1}$ and $\boldsymbol{Y}_{i3}$ mild positive correlation. The measurement times $t_{ij}$ are irregularly spaced and unsynchronized among different subjects. It was generated from $t_{ij}=\tilde{t}_{j}+e_{ij}$, where $\{ \tilde{t}_{j} \}$ is the scheduled measurement times from 0 to 12 years with 0.5 increment and $e_{ij}\sim Unif(-0.17,0.17)$. This setup corresponds to the practical situation where the subject had clinical visit within a two-month window around the scheduled visit times. For each simulation scenario, we used $500$ Monte Carlo repetitions and the sample size is $n=500$.

\section*{\textbf{Web Appendix B}: Simulation on Local Linear Estimation}

As explained in the Simulation section, the proposed landmark SDH model is a working model and it is therefore difficult to simulate data so that the model holds at all landmark times. This is a common feature of the landmark (or partly conditional) modeling approaches in general. In light of this difficulty, we resort to a simple albeit approximate approach to evaluating the quality of the proposed local linear estimation, at any landmark time $s$, as described below. 

We simulated a cross-sectional time-to-event data set at a given landmark $s$, e.g., $s=3$, which was treated as baseline for the purpose of this simulation. Scattered individual measurement times $\{ t_{ij} \}$ and the associated biomarker values $\boldsymbol{Y}_{i}(t_{ij})$ were simulated within a small neighborhood of $s$. The proposed landmark SDH model was used to generate independent competing risks data starting from each $t_{ij}$, following the simulation algorithm in Fine and Gray (1999). The log-SDH $\boldsymbol{\beta}(s)$ is assumed to be a quadratic function of $s$ (Web Figure 5). Note that this is not a really a landmark dataset because each subject only has one $t_{ij}$. Nonetheless, this dataset exactly satisfies the landmark SDH model so that we can use it to study the numerical performance of the proposed local linear estimation in a small neighborhood of $s$. Specifically, we evaluate the bias of estimating $\boldsymbol{\beta}(s)$ and the baseline CIF (Web Figure 6), $\pi_{0}(t^{*};s)=1-\textrm{exp}\Big(-\int_{0}^{t^{*}}\lambda_{10}(t,s)dt\Big)$, as well as the selection of the bandwidth. 

The results are presented in Web Figure 7. 
The three columns from left to right are the plots of the estimated log-SDH ratio, bias percentage, and mean squared error (MSE) against different bandwidths. The rows from top to bottom correspond to the three increasing sample sizes. For the plot of the log-SDH ratio (column 1), the mean estimated $\boldsymbol{\beta}(s)$ at $s=3$ over the Monte Carlo repetitions is close to the true value (red horizontal line) at small bandwidths (e.g. 0.3 and 0.5). With increased bandwidth, the estimator shows increasing downward bias. This is because the true $\boldsymbol{\beta}(s)$ function is concave (Web Figure 5), and the local linear fit underestimates it at the peak as the bandwidth increases. The empirical standard errors, shown in Web Figure 7 as the vertical whiskers, shrink with the increased bandwidth since more data points are included in the kernel estimation. From top to bottom, the empirical standard errors decrease when the sample size increases. Column 2 shows that the bias percentage generally increases with the bandwidth, except when the bandwidth is very small, in which case larger finite-sample bias may result due to very few data points available in the neighborhood defined by the bandwidth. In column 3, the U-shaped MSE curve is a demonstration of the typical bias-variance trade-off in kernel estimation. Overall, the percentage of absolute bias for the log-SDH ratio is very small, within 2\% for middle ranged bandwidths (the horizontal dashed line in column 2). The results from this simulation suggests that the proposed local linear estimation works as expected from typical local polynomial estimators. 

\section*{\textbf{Web Appendix C}: Table and Figures}

\begin{table}[htbp]
\centering
\caption*{\textbf{Web Table 1}: The predicted CIF at different landmark times $s$ and biomarker values $m$. The true conditional risk (True) were obtained empirically using the method described in Section 5. The average estimated CIF (EST), percent bias (\%Bias), empirical standard deviation (ESD), and mean-squared errors ($\times 1,000$) (MSE) are reported. Prediction horizon $\tau_1=3$. The result is based on 500 Monte Carlo repetitions. }
\begin{tabular}{ccccccc}
  \hline
&  & True & EST & \%Bias & ESD & MSE \\ 
  \hline
 & $m=0$ & 0.167 & 0.168 & 0.703 & 0.029 & 0.836 \\ 
$s=1$ & $m=2$ & 0.357 & 0.350 & -2.030 & 0.025 & 0.670 \\ 
 & $m=4$ & 0.610 & 0.639 & 4.734 & 0.052 & 3.583 \\ 
\hline
 & $m=0$ & 0.278 & 0.295 & 6.262 & 0.052 & 3.001 \\ 
$s=3$ & $m=2$ & 0.516 & 0.503 & -2.589 & 0.033 & 1.299 \\ 
 & $m=4$ & 0.729 & 0.755 & 3.463 & 0.053 & 3.419 \\ 
\hline
 & $m=0$ & 0.312 & 0.327 & 4.762 & 0.089 & 8.068 \\ 
$s=5$ & $m=2$ & 0.505 & 0.491 & -2.795 & 0.062 & 4.104 \\ 
 & $m=4$ & 0.681 & 0.691 & 1.407 & 0.064 & 4.160 \\ 
   \hline
\end{tabular}
  \label{tab:sim_predRisk}
\end{table}

\begin{figure}[ph]
  \centering{}\includegraphics[scale=0.8]{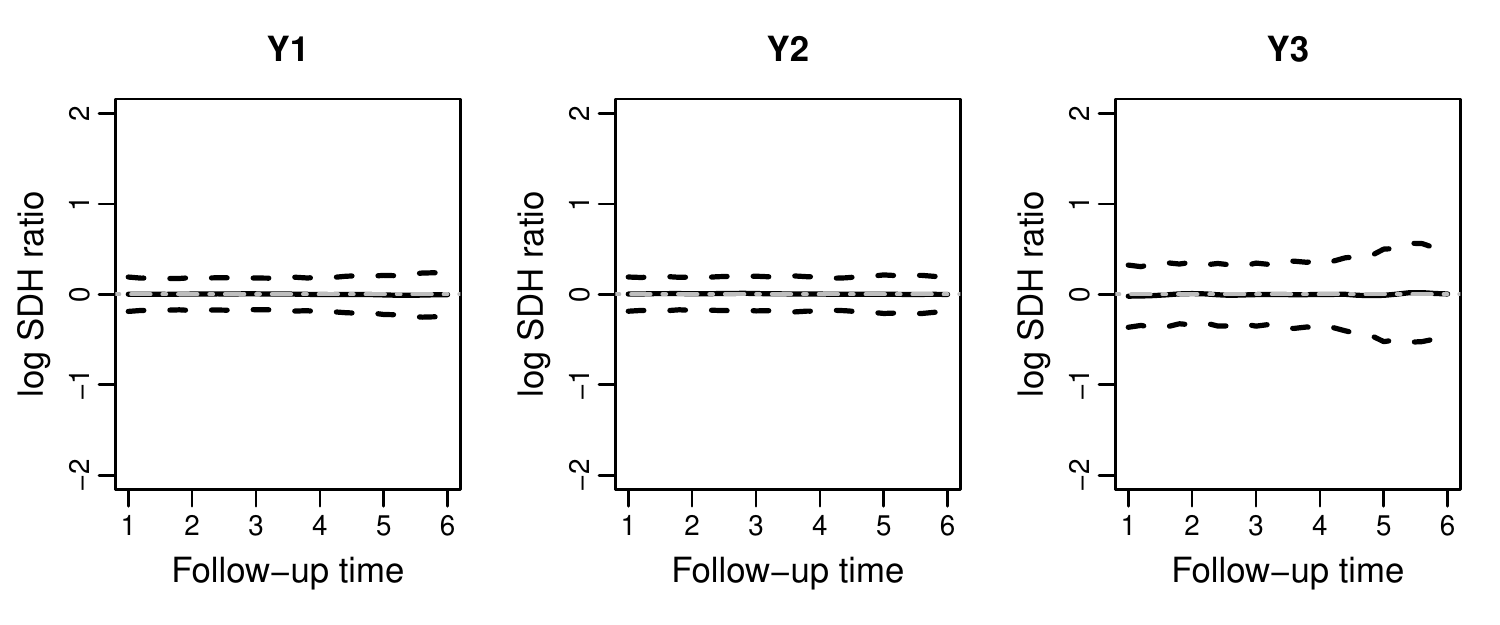}
  \centering{}\includegraphics[scale=0.8]{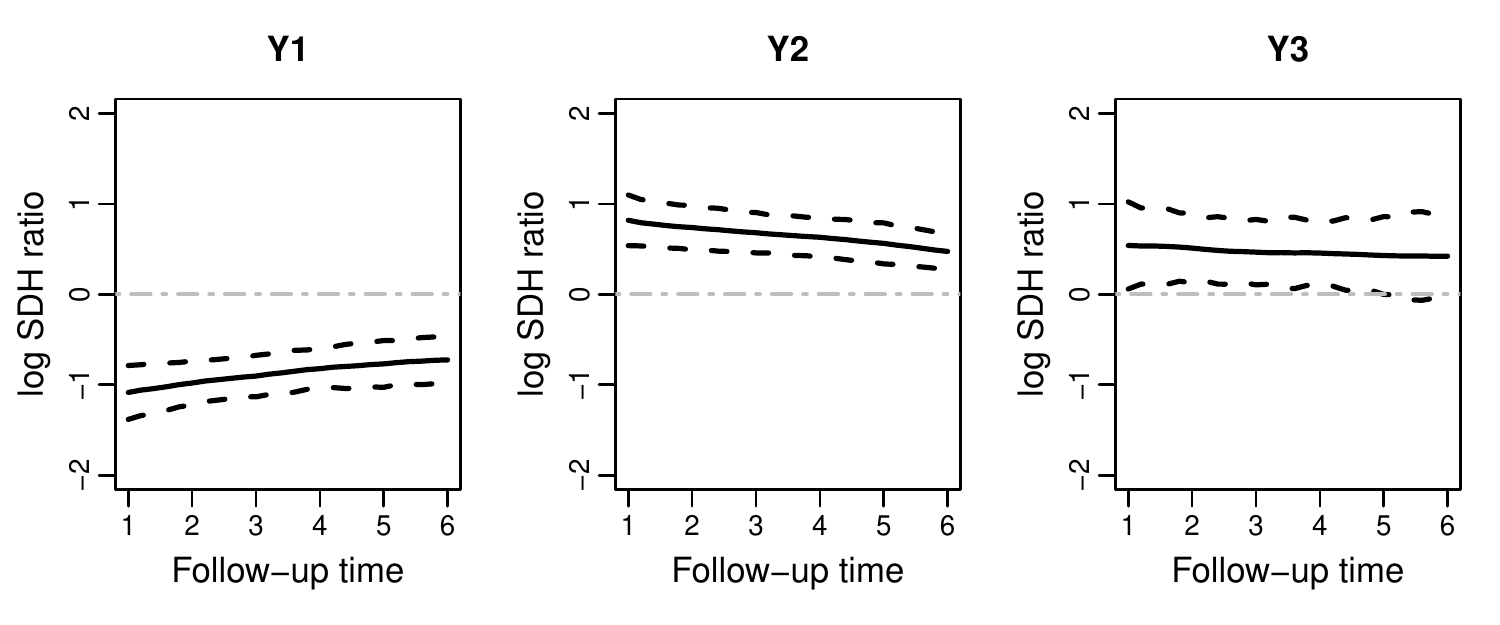}
  \caption*{\textbf{Web Figure 1}: Simulations with non-informative biomarker effect (upper panel) and informative biomarker effect (lower panel). The point estimator of log-SDH ratios corresponding to the three longitudinal biomarkers Y1, Y2 and Y3 (solid line) and their 95\% empirical confidence limits (dashed lines) are plotted over landmark time grids. The point estimator and the confidence limits are defined as the average, 2.5\% and 97.5\% quantiles of the point estimators from the Monte Carlo repetition.The horizontal dashed lines are the reference for zero effect. The estimated effects of the three biomarkers are close to zero when the biomarkers are non-informative and deviate from zero when the biomarkers are informative. Sample size $n=500$. } 
  \label{fig:simu_informative}
\end{figure}

\begin{figure}[ph]
  \centering
  \begin{subfigure}{1.0\linewidth}
    \centering
    \includegraphics[scale=0.7]{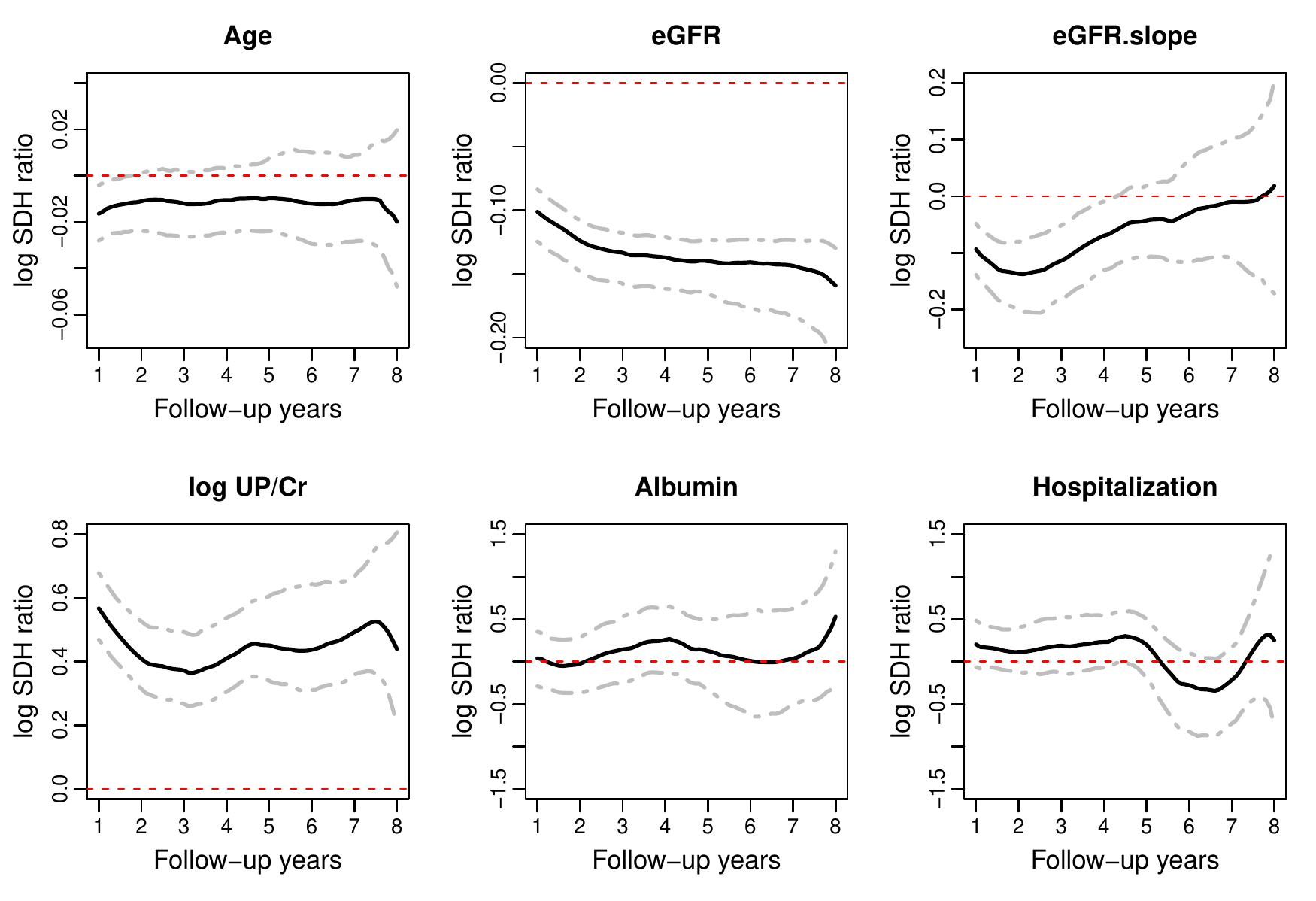}
    \caption{ESRD}
  \end{subfigure}
  
  \begin{subfigure}{1.0\linewidth}
    \centering
    \includegraphics[scale=0.7]{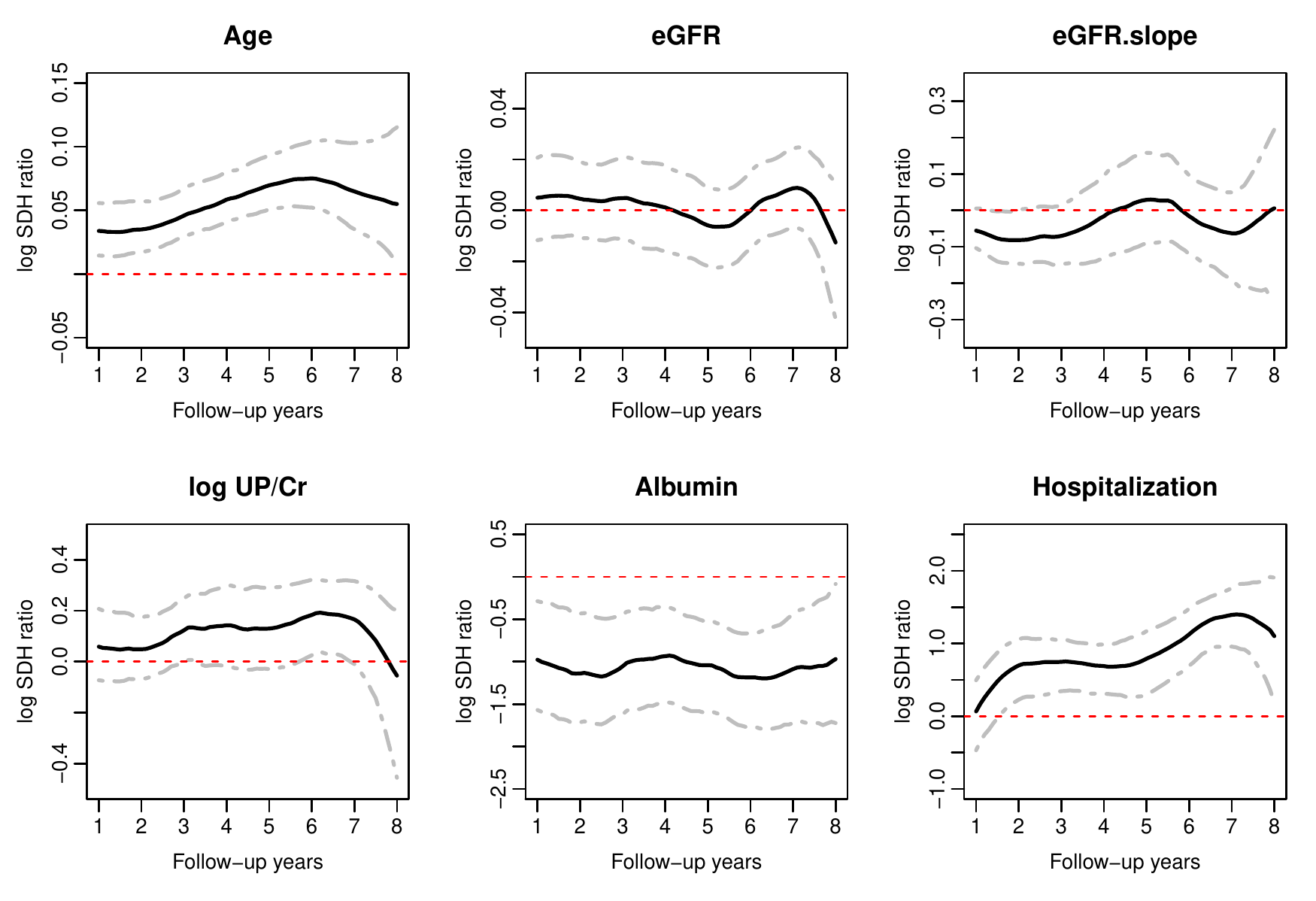}
    \caption{Death}
  \end{subfigure}
  \caption*{\textbf{Web Figure 2}: The time-varying log-SDH ratios of Age, eGFR, eGFR.slope, log UP/Cr, Albumin and history of hospitalization for the outcome of ESRD and death.
    Black solid curves are the time-varying log-SDH ratios and grey dashed
    curves are the 95\%  confidence intervals from bootstrap. Red
    dotted lines are the reference line of zero effect.}
  \label{fig:AASK_univ_betacurve}
\end{figure}

\begin{figure}[ph]
    \centering
    \begin{subfigure}{1.0\linewidth}
    \centering
        \includegraphics[scale=0.65]{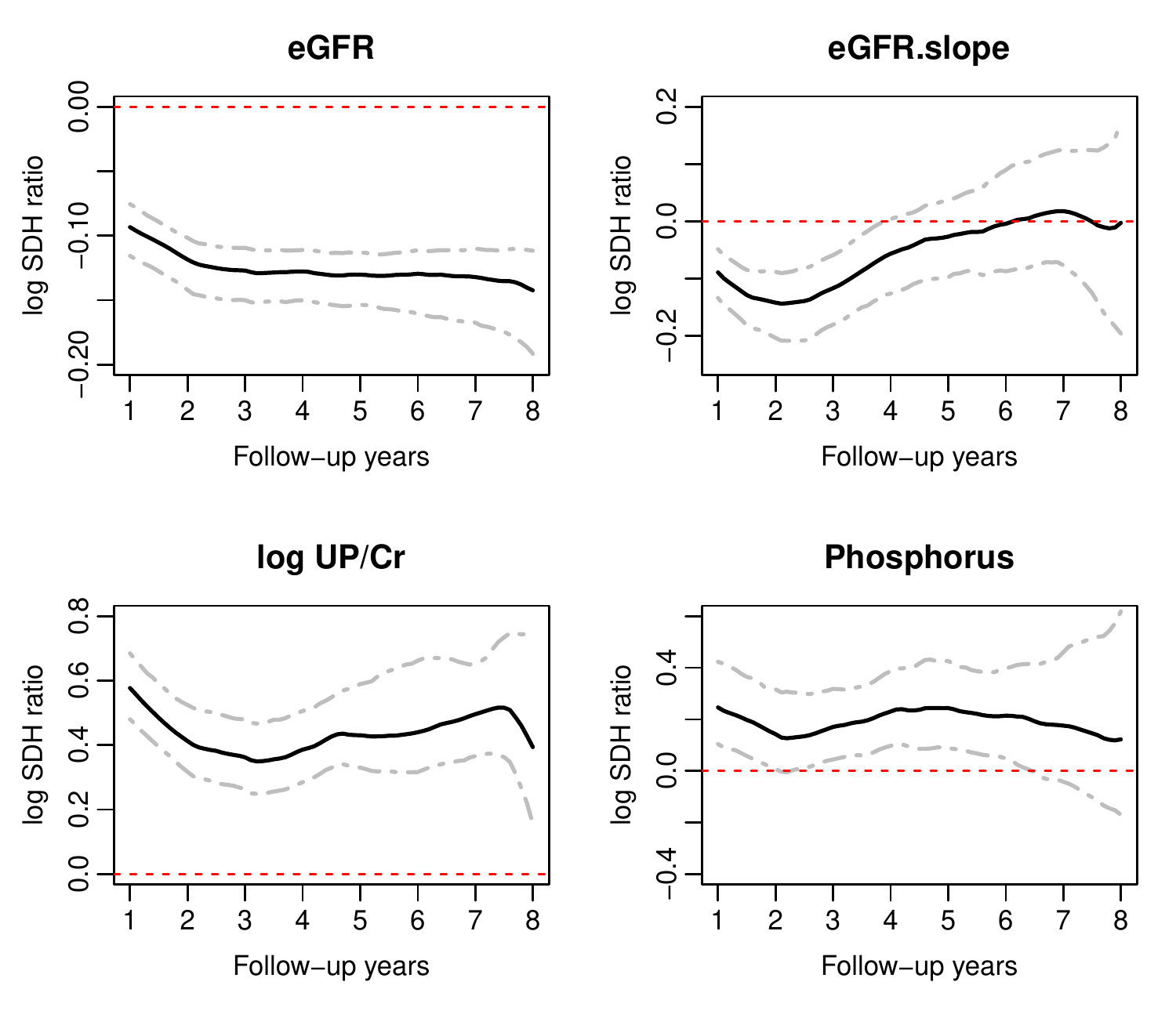}
        \caption{ESRD}
    \end{subfigure}
    
    \begin{subfigure}{1.0\linewidth}
    \centering
        \includegraphics[scale=0.65]{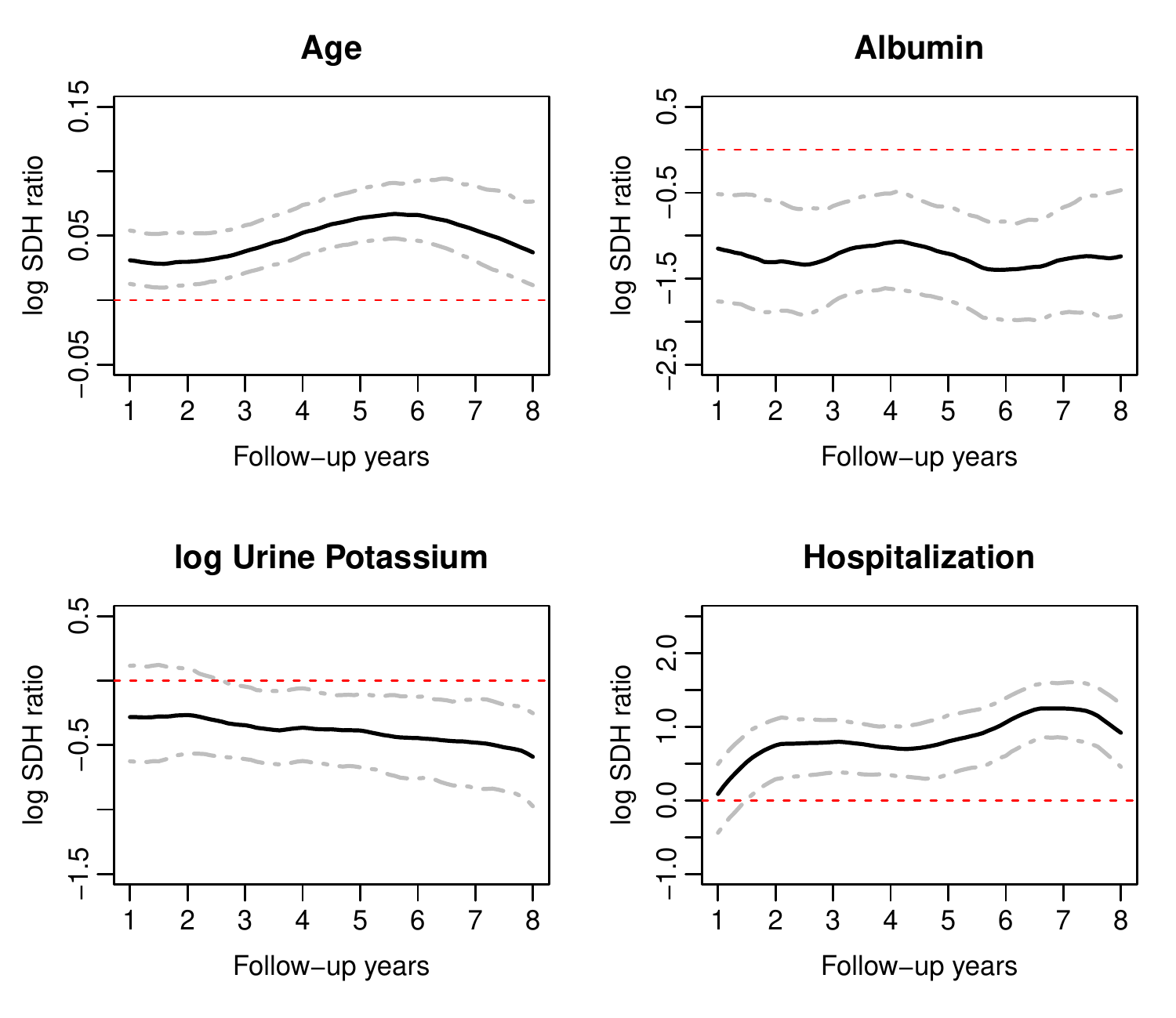}
        \caption{Death}
    \end{subfigure}
    \caption*{\textbf{Web Figure 3}: The time-varying log-SDH ratios of Age, eGFR, eGFR.slope, log UP/Cr and Phosphorus for the outcome of ESRD; and the time-varying log-SDH ratios of Age, Albumin, log Urine Potassium, and history of hospitalization  for the outcome of Death. 
    Black solid curves are the time-varying log-SDH ratios and grey dashed
    curves are the 95\% confidence intervals from bootstrap. Red
    dotted lines are the reference lines of zero effect.}
\end{figure}

\begin{figure}[ph]
  \centering{}\includegraphics[scale=1.0]{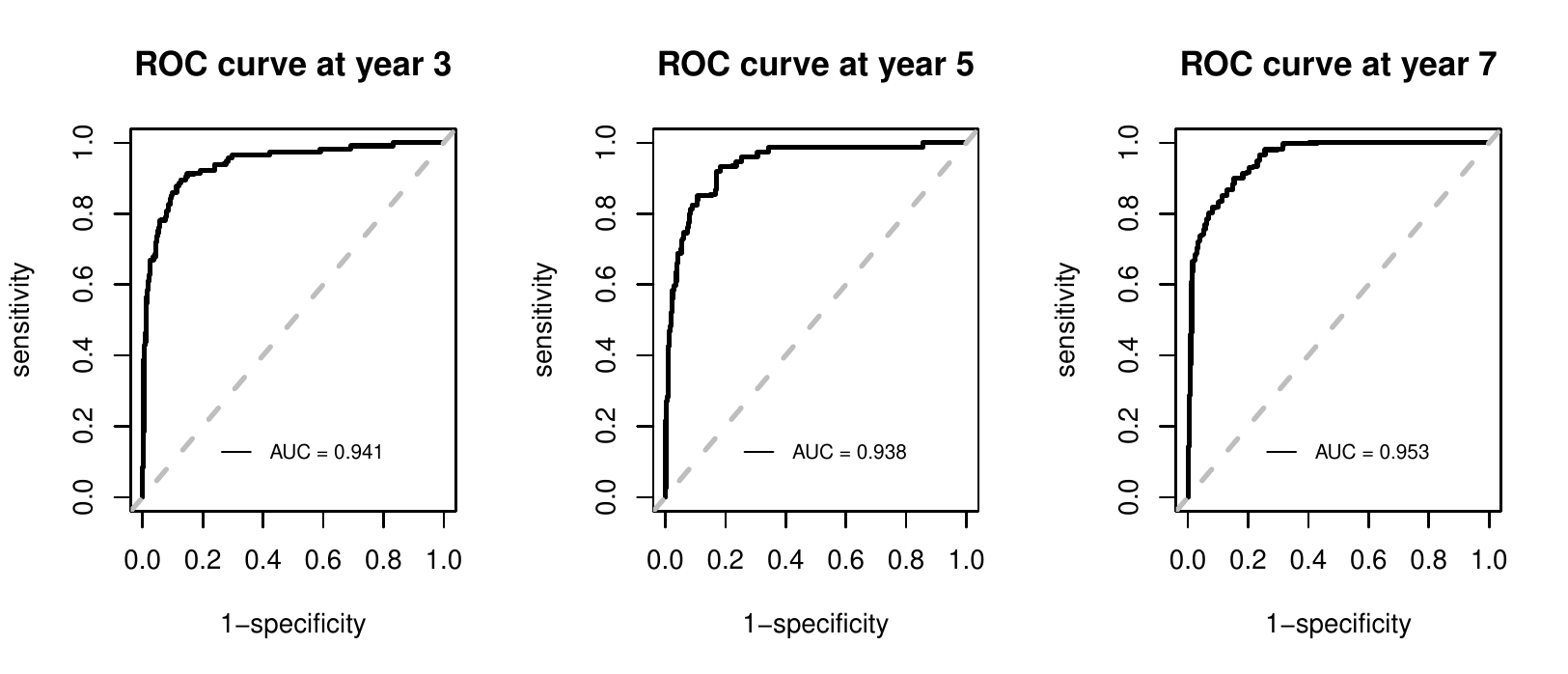}
  \caption*{\textbf{Web Figure 4}: The time-dependent ROC curves for predicting ESRD at landmark years
    3, 5, and 7, with prediction horizon $\tau_{1}=3$ years. The areas
    under the ROC curves (AUCs) are annotated on the plots. }
  \label{fig:AASK_ROC}
\end{figure}

\begin{figure}[ph]
  \begin{centering}
    \includegraphics[scale=0.6]{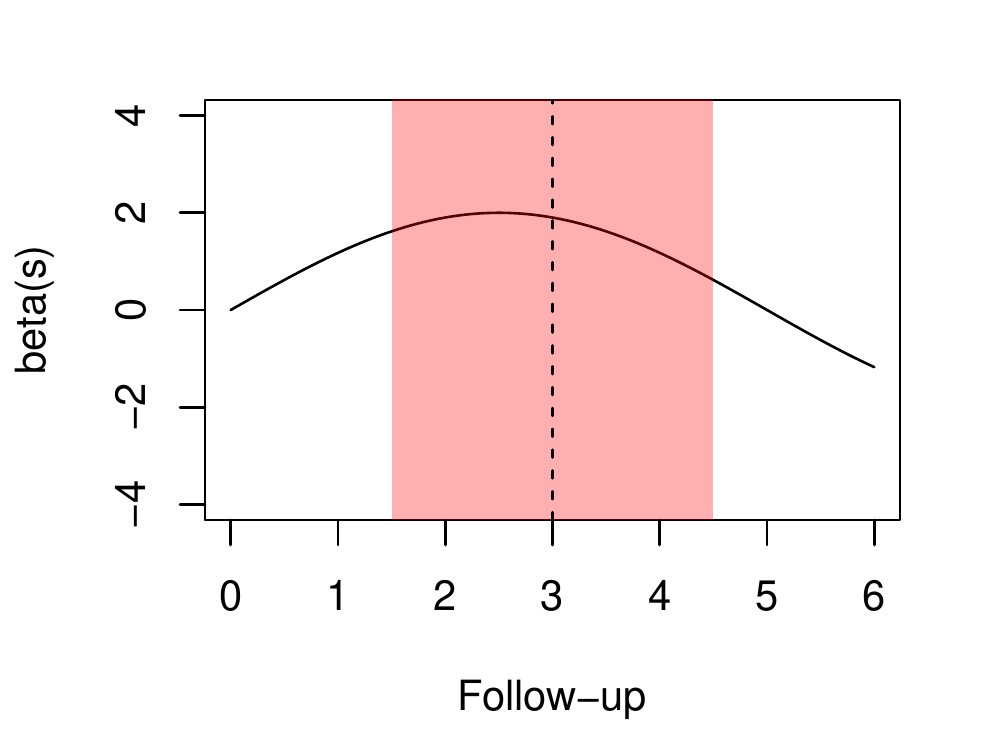}
    \par\end{centering}
  \caption*{\textbf{Web Figure 5}: Simulation setting for Web Appendix B. For each subject, one longitudinal biomarker value was simulated at a randomly picked time within a neighborhood (pink shaded interval) of landmark time $s=3$. The curve shows the shape of the coefficient function $\beta(s)$. }
  \label{fig:simu_true_beta_curve}
\end{figure}

\vspace{3cm} 

\begin{figure}[ph]
  \begin{centering}
    \includegraphics[scale=0.6]{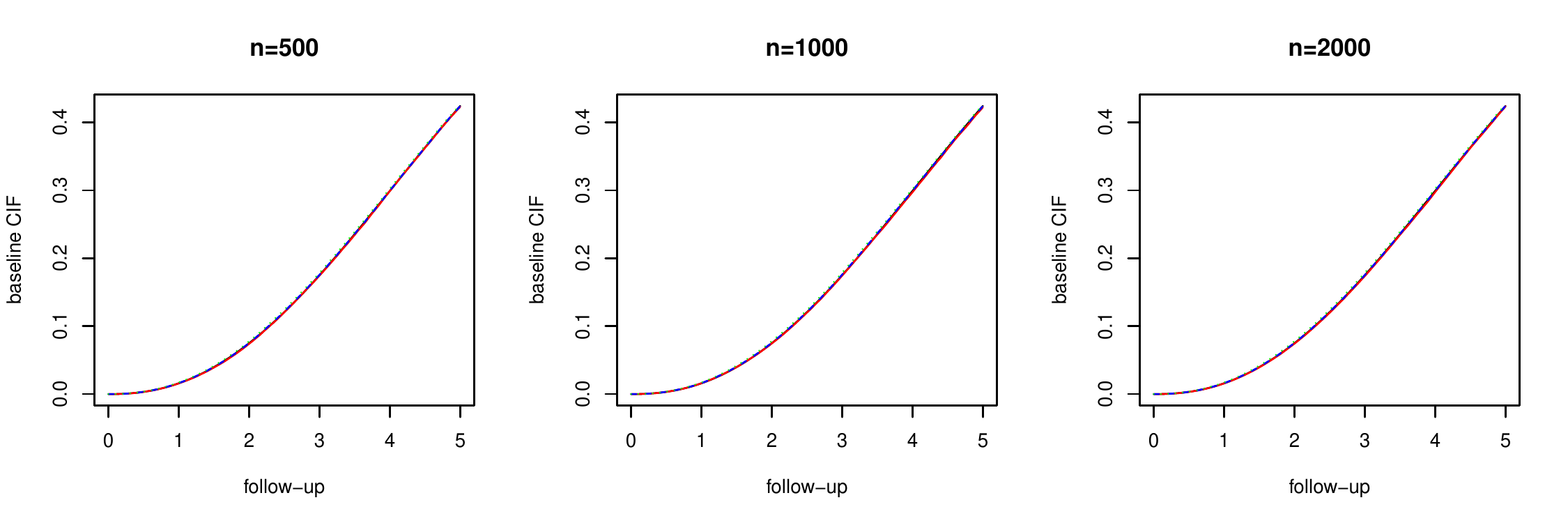}
    \par\end{centering}
  \caption*{\textbf{Web Figure 6}: Estimation of baseline CIF in the simulation of Web Appendix B. The sample sizes are 500, 1000 and 2000 respectively. True baseline CIF (red curve) and the average estimated CIF over the Monte Carlo repetitions nearly overlap. }
  \label{fig:nlin_cif}
\end{figure}

\begin{figure}[ht]
  \centering{}
  \includegraphics[scale=0.8]{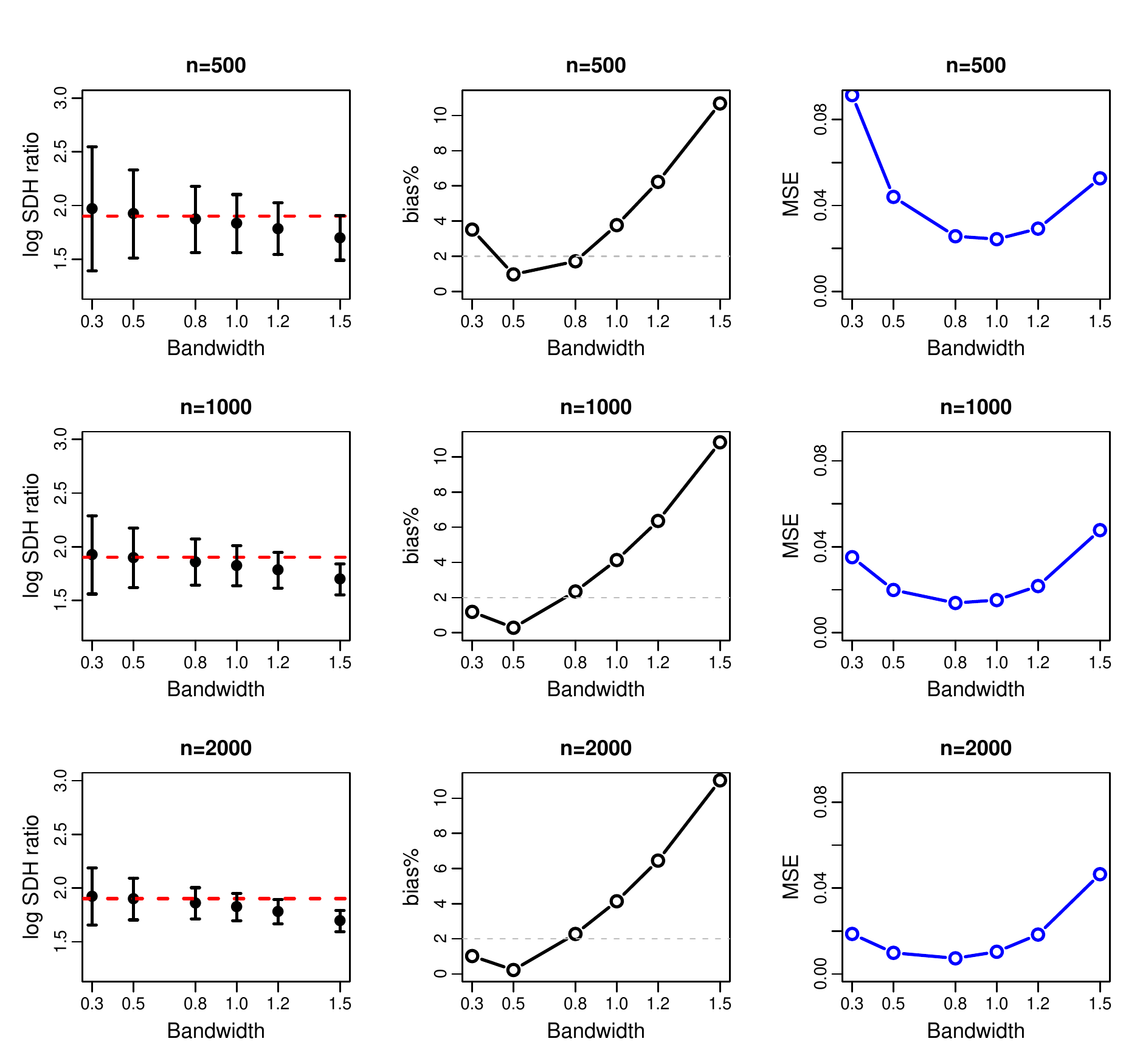}
  \label{fig_simu_kernel}
  \caption*{\textbf{Web Figure 7}: Simulation results for the finite sample performance of local linear estimation (Web Appendix B). The average estimated log-SDH ratio, absolute bias percentage and mean squared error (MSE) are plotted against the bandwidth on the horizontal axis. The sample size equals to 500, 1000, and 2000. The bias-variance trade-off and their relationship with the bandwidth resemble the typical behavior of local polynomial estimation. }
\end{figure}

%
%